\documentclass[aps]{revtex4}
\usepackage{amsmath,amssymb}
\usepackage{color,graphicx}
\usepackage{multirow}
\usepackage{amsmath,amssymb}
\usepackage{float}
\newcommand{\be}{\begin{equation}}
\newcommand{\ee}{\end{equation}}
\begin{document}
 \title{L\'{e}vy flights in steep potential wells: Langevin modeling versus direct  response to energy  landscapes}
 \author{Piotr Garbaczewski and Mariusz \.{Z}aba }
 \affiliation{Institute of Physics, University of Opole, 45-052 Opole, Poland}
 \date{\today }
 \begin{abstract}
We investigate  the   non-Langevin relative  of  the     L\'{e}vy-driven    Langevin  random system, under an assumption that both  systems   share a common   (asymptotic, stationary, steady-state)   target pdf.    The   relaxation to equilibrium  in the  fractional  Langevin-Fokker-Planck  scenario  results from an  impact of  confining conservative force fields  on the random motion.  A  non-Langevin  alternative  has  a   built-in  direct  response  of  jump intensities  to energy (potential) landscapes  in which the process takes place.   We revisit the problem of   L\'{e}vy flights   in superharmonic  potential wells,  with a  focus on  the  extremally steep  well  regime,   and  address the issue of   its   (spectral)  "closeness"   to  the     L\'{e}vy jump-type process confined  in  a finite enclosure  with impenetrable (in particular reflecting)   boundaries.  The  pertinent  random system   "in a box/interval"   might be  expected to have  a fractional Laplacian   with  suitable  boundary conditions  as a   legitimate motion generator. It is not the case. Another  problem is,  that in contrast to Dirichlet boundary problems, a concept of  reflecting   boundary  conditions    and  the    path-wise implementation  of the pertinent  random  process  in  the  vicinity,   or sharply at  reflecting  boundaries,    are  not unequivocally settled  for     L\'{e}vy  processes.  This ambiguity extends to   nonlocal analogs of Neumann conditions for fractional generators, which  do not comply with  the  traditional   path-wise  picture of  reflection  at  the impenetrable  boundary.
\end{abstract}
\maketitle

\section{Introduction}

The Eliazar-Klafter targeted stochasticity concept, together with that of the reverse engineering (reconstruction of the stochastic process once a target pdf is a priori given), has been originally devised for Lévy-driven Langevin systems. Its generalization, discussed in \cite{gar,gar1},   involves a non-Langevin alternative which associates with the Levy driver and the Langevin-induced  target pdf,  another (Feynman-Kac formula related) confinement mechanism for Lévy flights, based on a direct reponse to energy (potential) landscapes, instead of that to conservative forces.   

We revisit the problem of Lévy motion in  steep potential wells, analyzed  in terms of  a sequence of Fokker-Plack equations and their stationary solutions in  Refs.  \cite{dubkov,dubkov1}  and next  path-wise in Ref. \cite{dybiec}. Although we are ultimately interested in the  extremal steepness  regime, we need to mention that the  above  "sequential" strategy has  been  introduced and next  developed in a number of earlier publications \cite{chechkin,chechkin1,chechkin2},  and   summarized in a couple of review papers  \cite{chechkin3,chechkin4,chechkin5}, see also \cite{chechkin6}.

An  association   of  stationary probability density functions (pdfs),   arising in  the  sequential  superharmonic  approximation  (signatures of convergence),   with  steady-states of  L\'{e}vy flights in a confined domain   has been  reported.     The  "confined domain"  notion  has received an  explicit   interpretation of  the {\it infinitely deep potential well} enclosure,  \cite{denisov,zoia}, see also \cite{gar5,trap}.
This motivates our   investigation of   the     semigroup (Feynman-Kac) motion scenario, which actually provides a  non-Langevin alternative   to the Langevin-Fokker-Planck relaxation process of \cite{dubkov,dubkov1,dybiec}.   Our focus is on the possible asymptotic (growing steepness limit)  emergence of a   link with the problem of boundary data (Dirichlet versus Neumann, or absorbing versus reflecting) for the Lévy motion and its generator on the interval (or bounded/confined domain interpreted as the infinitely deep potential well), \cite{kulczycki}.

 One more important point should be raised. The infinite well enclosure  for the random motion,  and  the  limitation  of the latter  to  the interval interior  (impenetrability or inaccessibility of endpoints), are related to  the  notions of  absorbing (Dirichlet) and reflecting (Neumann-type) boundaries.  Interestingly, the discussion of \cite{dubkov,dubkov1,dybiec}  definitely   takes for granted the association of the  "confined domain"  (like e.g. the infnitely deep potential well)  boundaries with the reflection scenario  for random motion, which is not a must, c.f. \cite{trap,brownian,gar4}, see also \cite{kulczycki}. This point we shall briefly discuss in Section III of the present paper, where  the "confined domain" (and likewise the infinite well)  will  be    associated  with  Dirichlet boundary conditions.

As well, for the above mentioned "infinitely deep potential well problem",   no link  has been established  with the  Neumann fractional Laplacian     (whatever that is meant to be, neither with   any   convincing   form of the Neumann condition),   which is supposed to be a  valid   generator of  the  L\'{e}vy  process  in a bounded domain with reflecting boundaries.

This boundary issue can be consistently analyzed  by employing   the transformation  of  the  fractional Fokker-Planck equation to the    fractional Schr\"{o}dinger-type equation (hence to the fractional semigroup).
 It  is a properly tailored   version of  the  technical tool, often  used  in the study  case of the  standard (Brownian)  Fokker-Planck equation, but  seldom  addressed in the literature on  confined  L\'{e}vy processes.
 Since the superharmonic approximation seems  to provide a suggestive method to understand  what is   possibly  meant  by  the reflected L\'{e}vy process in the bounded domain (we restrict considerations to the interval on $R$), the  pertinent  transformation to the Schr\"{o}dinger-type  dynamics   should  in  principle   provide an approximation of this governed by the Neumann Laplacian and allow to identify features of the "spectral closeness" of  the pertinent operators (and fractional semigroups).

We shall investigate signatures of convergence for a sequence of   confined L\'{e}vy  processes on a line, in conservative
force fields   $\sim - x^{n-1}$ stemming from superharmonic potentials $U(x)\sim x^n, x^n/n, n x^n$ with  $n = 2m \geq  2$.
This is paralleled by a transformation of  the related  fractional   Fokker-Planck operator  $L^*= - |\Delta |^{\alpha /2} - \nabla [b(x)\cdot ]$   into the    fractional  Schr\"{o}dinger-type operator $\hat{H} = |\Delta |^{\alpha /2} + {\cal{V}}$, whose potential function is inferred from the knowledge of  the square root of the stationary probability density $\rho _*(x)$ of the corresponding Markov process, according to ${\cal{V}}= - \rho _*^{-1/2}   |\Delta |^{\alpha /2} \rho _*^{1/2} $. The  pertinent $\rho _*^{1/2}(x)$  actually is  the  $L^2(R)$-normalized  ground state  function of $\hat{H}$.

The transformation of the  Fokker-Planck operator (and likewise of  the adjoint  random motion generator)    into the Schr\"{o}dinger one  is a celebrated  tool in the study of  Brownian  relaxation processes. Recently, we  have  discussed  at length  various aspects of  superharmonic approximations of the Brownian motion in the interval  (and links of the latter problem with that of  the  "infinitely deep potential well").  The  Brownian route has been chosen as a playground for checking jeopardies and possible inadequacies of the (Schr\"{o}dinger)  transformation methodology \cite{brownian}, prior to passing to an analysis  of superharmonic approximations  of  L\'evy flights "in the infinitely deep well" (or interval),  and  technically  more involved issue of reflected L\'{e}vy flights in the  "box"  enclosure with impenetrable walls/barriers,  along the lines indicated in Ref.  \cite{smol}.

  Our discussion goes far beyond   the technical  (transformation proper) features.  We deal here  with  physically different mechanisms for   the response of  the  symmetric  stable noise in one space dimension  to  external perturbations. To be  considered   as alternative response (and motion)    scenarios:   (i) set   by conservative    force  potentials (motion in energy landscapes),    or   (ii) set directly   by    force fields.  For both types of perturbations we shall preselect the L\'{e}vy  driver and the  stationary target  pdf  (common for both Langevin and non-Langevin   motion scenarios), i.e. a  probability  density  function $\rho _*(x)$ to which the  random  process asymptotically  relaxes, once  started   with a suitable initial pdf $\rho _0(x)$:  $\rho _0(x) \rightarrow  \rho (x,t) \rightarrow \rho _*(x)$.

   The Langevin  approach  stems from   so-called targeted stochasticity  concept,  addressing the issue of an  attainability of equilibria (generically not of the Boltzmann-type)   for  L\'{e}vy-driven   Langevin systems,  \cite{eliazar}, see also \cite{gar,gar1,gar2}.  A  related idea  of  reverse engineering,  refers to  a  reconstruction  (designing) of   a L\'{e}vy-Langevin system,  that would    yield (relax to)    a pre-defined  target pdf.  The   pertinent random motion is known not to obey the detailed balance condition, \cite{balance,balance1}, while the non-Langevin dynamics by construction does.

  The non-Langevin approach, may be interpreted as  an alternative   version of the reverse engineering procedure  (stochastic process reconstruction). It has roots in  Ref. \cite{sokolov},  see also \cite{gar,gar1,gar2}, and involves  the "potential landscape" idea of Ref. \cite{kaleta}.  Given a stationary pdf   $\rho _*(x)$  and the L\'{e}vy driver, one  specifies the semigroup dynamics  whose  generator  (fractional Laplacian plus a suitable  potential function)   has that  pdf  square root  $\rho _*^{1/2}(x)$  as the  positive-definite  ground state function, \cite{vilela,faris}.  The  semigroup dynamics can be elevated to the fully-fledged  stochastic process,  governing the relaxation of  a suitable $\rho (x,t)$   to the pre-defined $\rho _*(x)$, while maintaining the detailed balance condition.

 \subsection{L\'{e}vy driver.}

  Let us set the basic framework and the notation, to keep it   uniform throughout the paper.
 A characteristic function of a random variable $X$  completely
determines a probability distribution of that variable. If this
distribution admits a density $\rho(x)$, we can write $<\exp(ipX)> =
\int_R \rho (x) \exp(ipx) dx$ which, for infinitely divisible
probability laws,  gives rise to  the famous L\'{e}vy-Khintchine
formula.   From now  on, we concentrate on the integral part of the
L\'{e}vy-Khintchine formula, which is responsible for arbitrary
stochastic jump features:
\begin{equation}
F(p) = - \int_{-\infty }^{+\infty } \left[\exp(ipy) - 1 -
\frac{ipy}{ {1+y^2}}\right] \nu (dy),
\end{equation}
where $\nu (dy)$ stands for the appropriate L\'{e}vy measure. The
corresponding non-Gaussian Markov process is characterized by
$<\exp(ipX_t)>= \exp[-t F(p)]$ and, upon setting  $\hat{p} = - i\nabla $
 instead of $p$,  yields an operator $F(\hat{p})$ which we interpret  as  the free Schr\"{o}dindger-type  Hamiltonian (for clarity of discussion, all dimensional constants  generally  are  scaled away, note e.g. that in the Gaussian case $F(\hat{p}) = - \Delta $).

We restrict further  considerations to non-Gaussian random variables
whose probability densities are centered and symmetric, e.g.  a
subclass of   $\alpha $-stable distributions admitting a straightforward definition of the fractional  noise  generator
\begin{equation}
F(p) =   |p|^{\alpha  }   \to F(\hat{p})  \doteq
|\Delta |^{\alpha  /2} =  (- \Delta )^{\alpha /2} .
\end{equation}
We indicate that the adopted definition  of the fractional Laplacian  coincides with   the negative of  a suitable Riesz fractional derivative   $\partial ^{\alpha } / \partial |x|^{\alpha}\equiv  \Delta ^{\alpha /2}$,
 e.g.    $ (- \Delta )^{\alpha /2}= - \partial ^{\alpha } / \partial |x|^{\alpha  }$.

 In the above,   $0<\alpha  <2$  stands for a  stability index of the L\'{e}vy noise and the related stochastic  process.  The   fractional  Laplacian    is a non-local pseudo-differential operator, by construction
nonnegative   and self-adjoint on a properly tailored domain.
The  induced   jump-type   dynamics is interpreted in terms of
L\'{e}vy flights.   In particular   $\alpha  =1$  refers to the Cauchy process,  with the generator (for the record we list  varied, commonly   used notational conventions )   $ F(\hat{p}) =  |\nabla |=    |\Delta |^{1/2}  =   (-\Delta )^{1/2} \equiv \sqrt{- \Delta }$.

The pseudo-differential Fokker-Planck equation  derives from the fractional  semigroup $\exp(- t |\Delta |^{\alpha
/2})$  and   reads  $\partial _t \rho  = -  |\Delta |^{\alpha  /2} \rho $.  That to be compared
 with the   "normal" Fokker-Planck equation for a freely diffusing  particle  (Wiener   noise, with noise intensity $D=1$)  $\partial _t \rho =  \Delta \rho $,  deriving from the semigroup  $\exp (t \Delta )$.

An explicit  integral  form of the a pseudo-differential operator $|\Delta |^{\alpha /2}$, follows  from (1) and (2):
 \begin{equation}
(|\Delta |^{\alpha  /2} f)(x)\, =\, - \int_R [f(x+y) - f(x) - {{y\, \nabla f(x)}
\over {1+y^2}}]\, \nu _{\alpha  }(dy).
\end{equation}
 This expression can  greatly simplified, in view of
the properties of the  L\'{e}vy measure $\nu _{\mu }(dx)$.  Namely, remembering that we overcome a
 singularity at $0$  by means of the  Cauchy   principal value  of the  integral, we may replace (3) by
\be
 |\Delta |^{\alpha /2} f(x)\,   = (-\Delta )^{\alpha /2}f (x)   =\, - \int_R  [f(x+y) - f(x) ] \nu _{\alpha }(dy) \, .
\ee
By changing  an integration variable $y$  to $z=x+y$ and employing  a direct connection  with the Riesz fractional derivative of the  order  $\alpha $,  we  arrive at
\begin{equation}
  |\Delta |^{\alpha  /2} f(x)\, =\, -  \mathcal{A} _{\alpha }  \int_R  {\frac{f(z)- f(x)}{|z-x|^{1+\alpha }}}\,
 dz.
 \end{equation}
 where ${\cal{A}}_{\alpha }= \pi ^{-1} \Gamma (\alpha  +1) sin(\pi \alpha /2)$.
The case of $\alpha  =1$  refers to the  Cauchy driver, with  $\nu _{1} (y)= 1/\pi y^2$.

We point out that,  the   evaluation of   the singular integral in the definitions (4) and  (5), needs some care. In Eq. (3) the problem is bypassed by means of the counter-term.   An alternative definition:
\be
 |\Delta |^{\alpha  /2} f(x)\,= (-\Delta)^{\alpha /2}f(x)=  {\frac{\mathcal{A}_{\alpha}}{2}} \int _{R}
\frac{2f(x) - f(x+y) -f(x-y)}{|y|^{1+\alpha }}\, dy,
\ee
if employed in suitable function spaces,  is by construction free of   singularities  and happens to be more amenable to computational procedures,  \cite{kwasnicki,kwa}.

\subsection{Langevin-induced fractional Fokker-Planck equation and motion generators.}

In case of jump-type (L\'{e}vy) processes,  a response   of noise  to
 conservative force fields  may be  quantified  by mimicking  the Brownian pattern.
A  popular  reasoning, \cite{fogedby},   employs   a  (formal)
Langevin-type  equation   $\dot{x}= b(x)  + B^{\alpha  }(t)$    with  a deterministic term $b(x)$  (a gradient function $b \sim  -\nabla U$, presumed to  encode  Newtonian force fields)  and   the  additive L\'{e}vy "white noise" term.   This leads to   a fractional Fokker-Planck
equation (\cite{fogedby}, compare e.g.  also \cite{olk})  governing the time evolution  of the probability density function (pdf) $\rho (x,t)$  of the process:
\begin{equation}
 \partial _t\rho = -\nabla (b\cdot \rho ) -  |\Delta |^{\alpha /2}\rho \, .
 \end{equation}
We emphasize a difference in sign in the
second term, if compared with Eq. (4) of Ref. \cite{fogedby}.  There, the minus sign is  absorbed in the adopted definition of the (Riesz) fractional derivative. Apart from the formal resemblance of operator symbols, we do not directly employ fractional derivatives in our discussion.

Let us assume that the fractional Fokker-Planck equation (7)
admits   a   stationary  solution $\rho _*(x)$, which is an asymptotic target $\rho (x,t) \rightarrow \rho _*(x)$     of    the  relaxation process.  Then,  a functional form of the  time-independent  drift $b(x)$
 can be reconstructed by means of  an indefinite integral
\begin{equation}
b(x)= -    {\frac{\int  dx \,  |\Delta |^{\alpha  /2}\rho _*(x)}{\rho _*(x)}} \, .
\end{equation}
This is an ingredient  of the reverse engineering procedure, \cite{eliazar,gar,gar1},  of  reconstructing the random motion from the prescribed  stationary (target) pdf, once the   L\'{e}vy driver is  pre-selected.

Anticipating further discussion, let us introduce some elements of   the standard stochastic inventory. Let $p(y,s,x,t)= p(t-s,y,x), t>s\geq 0$    be the time-homogeneous transition density of  the relaxation process (7), e.g.
\be
\rho (x,t)  = \int_R p(y,s,x,t) \rho (y,s) dy
\ee

We shall pass to the notation $p(t,y,x)$  to enable a direct comparison with the exemplary   construction of the   Ornstein-Uhlenbeck-Cauchy process in  Ref. \cite{olk}.  Leaving aside unnecessary here mathematical details, we recall that the  pertinent stochastic (jump-type) process is generated by  the semigroup $T_t$, transforming  continuous functions on $R$ (of the class $C_0(R)$) as follows:
\be
T_t f(x)  = \int_R p(t,x,y) f(y)  = f(x,t)
\ee
In the mathematically oriented  literature it is common to  interpret  $p(t,x,y)dy$ as a probability of getting from $x=X(0)$ to the   (infinitesimal) vicinity of a point $y = X(t)$.   Accordingly,  $T_t f(x)  = E_x[f(X_t)]$  stands for a  conditional expectation value of an "observable"  $f(x)$, evaluated  over   endpoints $X(t)=y$  of   sample paths  started  from   $x =X(0)$  and terminated at time $t$.

The semigroup   $T_t = \exp(tL)$    has  a   generator
\be
 L f(x)= \lim _{t\downarrow 0}  {\frac{1}t} \left[   \int p(t,x,y) f(y)dy  - f(x)\right],
\ee
whose generic form reads
\be
L =  - |\Delta |^{\alpha /2}   + b \nabla .
\ee
(We note its  formal    resemblance to the generator of the  standard  diffusion process (e.g. Brownian motion)   $L_B = \Delta  + b\nabla $.)   Clearly, we have $\partial _t f(x,t) = L\, f(x,t)$.

The time evolution of  probability measures and associated probability density functions $\rho(x,t)$ is  governed by the  adjoint semigroup  $T^*_t = \exp (L^*t)$:
\be
T_t^* \rho (x,t) = \int p (t,y,x) \rho (y) dy.
\ee
Accordingly, we have  $\partial _t \rho (x,t) = L^* f(x,t)$, where
\be
L^*  = - |\Delta |^{\alpha /2}  - \nabla (b\,  \cdot )
\ee
comes from
\be
L^* \rho (x)= \lim _{t\downarrow 0}  {\frac{1}t} \left[   \int p(t,y,x) \rho (y) dy  -  \rho (x)\right].
\ee
See e.g. Refs.\cite{fogedby,olk} for exemplary calculations.

  We point out  that transition pdfs  in general  are not symmetric  functions   of  spatial variables:  $p(t,x,y)\neq p(t,y,x)$. The order of variables  clearly   identifies  the starting point (predecessor) and the terminal point (successor) for the  stochastic  process  (bridge) connecting these points   in  the time  interval of length   $t$.
In the mathematically oriented literature the pertinent symmetry is routinely restored, by  passing from Lebesgue to weighted integration measures, see for example \cite{kaleta,kaleta1,lorinczi}.

\section{Non-Langevin  approach.}

\subsection{Schr\"{o}dinger's interpolation problem.}

We are inspired by  apparent affinities between structural properties of  probabilistic solutions of the so-called Schr\"{o}dinger boundary data problem, \cite{zambrini,gar3},     and   the  current research on   conditioning of  Markovian stochastic processes, of diffusion and jump-type, \cite{kaleta,lorinczi,gar4,gar5,gar6}, see also \cite{gar,gar1,gar2}.
The Schr\"{o}dinger boundary data and interpolation problem  is known to provide  a unique Markovian interpolation between any two strictly positive probability densities,  designed to form the input–output statistics data  for a  random  process bound to   run  in a  finite (observation)   time interval.

The key input, if one attempts to reconstruct  the   pertinent Markovian dynamics, is to select the jointly continuous in space variables, positive and contractive semigroup kernel. Its  choice  is arbitrary,
except for the strict positivity  (not a must, but we keep this restriction in the present paper) and continuity demand. It is thus rather natural to ask for the most general stochastic interpolation, that is admitted under the above premises and the involved semigroups  may refer not merely to diffusion scenarios of motion, but  more generally to a broad family of non-Gaussian (specifically jump-type) processes.

The semigroup dynamics in question we  infer  from the  classic notion of  the  Schr\"{o}dinger  semigroup   $\exp(-t \hat{H})$, with a proviso that the semigroup generator $\hat{H}$  actually  stands  for  a  legitimate
(up to scaled away physical constants)  Hamiltonian operator,  incorporating   additive perturbations (by suitable potential functions) of either the traditional minus Laplacian, or   the fractional Laplacian of the preceding subsections.

  We are  interested in  Schr\"{o}dinger-type  operators of the form  $\hat{H}_{\alpha } =  (-\Delta )^{\alpha /2} + V(x)$, where   $\hat{H}_2 \equiv  - \Delta +  V(x)$,   and the  semigroups in question  appear as members of the $\alpha $-family $\exp (- t\hat{H}_{\alpha })$, with $0< \alpha \leq 2$.
  Although,  in our discussion the  Schr\"{o}dinger interpolation is restricted to a finite time interval $ t\in [0,T]$, this restriction may be relaxed, once the solution (e.g. transition probability of the process) is in hands.

Roughly, the essence of the Schr\"{o}dinger  boundary data problem,  \cite{zambrini},  goes as follows.
We consider Markovian propagation scenarios,  with the input - output statistics data  provided in terms of two strictly positive boundary densities $\rho (x,0)$ and  $\rho (x,T)$,   $T>0$, \cite{zambrini,gar3}, that may be constrained to (integrated over) some Borel sets $A$ and $B$  contained in $R$.
 We interpret   $\rho _0 (A)$ and $\rho _T(B)$   as   boundary data for a certain bivariate probability measure  $m(A,B)$.   Assume that the pertinent measure admits a  transition probability density
   \be
 m(x,y)= f(x) k(x,0,y,T) g(y)
 \ee
 with marginals  $\int _R m(x,y)dy = \rho (x,0)$ and $\int _R  m(x,y)dx  = \rho (y,T)$   presumed to be associated with   a certain dynamical process bound to run in a time interval $[0,T]$.

 Here,  $f (x)$ and $g(y)$  are the a priori unknown   strictly positive  functions,   that need to be deduced  from  the imposed boundary data (i.e. marginals that are presumed to be known a priori).  To this end,
   we  should select  any strictly positive, jointly continuous in space variables   kernel function $k(x,0,y,T)$.  We   impose a restriction that $k(x,0,y,T)$ represents a certain strongly continuous dynamical semigroup kernel      $k(y,s,x,t),\,   0\leq s<t \leq T$,   while specified at the time interval  $[0,T]$  borders. This assumption will secure  the Markov property of the sought for stochastic process.
Actually, we shall consider time homogeneous processes generated by the semigroup $\exp[- (t-s)\hat{H}_{\alpha })$, with a  kernel $k(t-s,y,x)$.

Under those circumstances,  \cite{zambrini,gar3},  once we define functions
\be
\theta (x,t) = \{ \exp[-(T-t)\hat{H}_{\alpha }]\, g \} (x)  = \int  k(x,t,y,T) g(y) dy
 \ee
 and
 \be
  \theta _*(y,t)   =    \{ \exp( -t\hat{H}_{\alpha }) \, g\}(y) =  \int k(x,0,y,t) f(x) dx
\ee
one can demonstrate  the existence of  a transition  probability  density  (note that even if $k(t,y,x)=k(t,x,y)$ the symmetry property is not respected by $p(t,x,y)$  in below)
\be
p(y,s,x,t) = k(y,s,x,t) {\frac{\theta (x,t)}{\theta (y,s)}},
\ee
which implements a Markovian propagation of the probability density
\be
\rho (x,t) =  \theta (x,t) \theta _*(x,t),
\ee
according to  the pattern
\be
 \rho (x,t) = \int  p(y,s,x,t) \rho (y,s) dy= \theta (x,t) \int k(y,s,x,t) \theta _* (y,s) dy = \theta (x,t) \theta _*(x,t),
\ee
 providing an interpolation between the prescribed boundary data in the time  interval $[0,T]$.

Here we note the exploitation of the semigroup property, \cite{zambrini}, in propagation formulas (17).
Namely,    for $0<s<t<T$,   we have
\be
\theta (x,s)=  \int  k(x,s,y,T)g(y) dy = \int \int k(x,s,z,t) k(z,t,y,T) g(y) dy dz =
\int k(x,s,z,t) \theta (z,t) dz
\ee
and
\be
\theta _*(y,t) = \int k(x,0,y,t)f(x)dx = \int \int k(x,0,z,s) k(z,s,y,t)f(x) dx dz =
 \int  k(z,s,y,t) \theta _*(z,s) dz
\ee

For a given semigroup    $\exp (-t\hat{H}_{\alpha })$ which is characterized by its Hamiltonian  generator $\hat{H}_{\alpha }$,  the kernel $\exp(-t\hat{H}_{\alpha })(y,x) = k(t,y,x)$ and the emerging transition probability density  $p(t,y,x)$ of the time homogeneous stochastic process  are unique in view of the
uniqueness of solutions $f (x)$ and $g(y)$ of the Schr\"{o}dinger boundary data problem, \cite{zambrini}.
  In  the  case of Markov processes, the knowledge of the transition probability density $p(y,s,x,t)$ (here $p(t-s,y,x)$) for all intermediate times $0\leq s<t \leq T$ suffices for the derivation of all other relevant characteristics of random motion.

Further exploiting the Schr\"{o}dinger semigroup lore, and their  Hamiltonian  generators,   we can  write evolution equations  for functions (17) in a  form  displaying an intimate link with   Schr\"{o}dinger-type  equations. Namely, while in the interval $[0,T]$, \cite{gar6}, se e.g (17) and (18),   we have   $\partial _t \theta _*=  -\hat{H} \theta _* $  and $ \partial _t \theta =  \hat{H} \theta $, where $\hat{H} = \hat {H}_{\alpha } =  |\Delta |^{\alpha /2} + V$.   Accordingly, \cite{gar3}:
\be
\partial _t\theta _*= - |\Delta |^{\alpha /2} \theta _*  - V \theta _*
\ee
and
\be
\partial _t\theta = |\Delta |^{\alpha /2} \theta  +  V \theta  .
\ee
For comparison, we indicate that the Brownian  version of Eqs. (24)  and (25) would have  the form  (up to scaled away physical constants)  $\partial _t\theta _*=  \Delta  \theta _*  -  V \theta _*$ and
 $\partial _t\theta = -\Delta  \theta  +  V \theta $ respectively.\\

{\bf Remark 1:}  At this point  let us recall  basic (Brownian)  intuitions that underly the the implicit  path integral formalism for L\'{e}vy flights.  Namely,  operators of the form  (the diffusion coefficient $D$ is scaled away, typically a dimensionless form $D=1/2$ or $D=1$ is used to simplify calculations)   $\hat{H}= - \Delta +  V \geq 0$  with $V \geq 0$  give rise to transition kernels of diffusion -type   Markovian processes with killing (absorption), whose rate is determined  by the value of $V(x)$ at $x\in R$. This  interpretation stems from the celebrated Feynman-Kac (path integration) formula, which assigns to $\exp(-\hat{H}t)$
 the positive  integral kernel
$$
  k(x,s,y,t)  =  [\exp(- (t-s)(-\Delta  + V)](y,x)  =
       \int   \exp[-\int_s^t V(\omega(\tau  )) d\tau ]\,   d\mu _{s,y,x,t}(\omega)
  $$
 In terms of Wiener paths the kernel is constructed as a path integral over paths  which  get killed at a point $X_t=x$,   with an extinction  probability $V(x)dt$     in the time interval$(t,t+dt)$.     The killed path is henceforth removed from the ensemble of on-going Wiener paths.   The exponential factor $\exp[-\int_s^t V(\omega (\tau ))d\tau ]$  is here responsible for a proper redistribution of Wiener paths, so that  the evolution rule
$$
f(x,t) =(\exp (-t\hat{H})]f)(x)  = \int _{R^n} k(x,0;y,t) f(y) dy   =  E^x [f(X_t)]  =  E^x[f(X_t) \exp (-\int_0^t  V(X_s )ds )],$$
 with $\hat{H}= -\Delta  +  V$,  is well defined as an expectation value of the killed process  $X(t)$, started at time zero, at $x\in R$.

  Anticipating further discussion, we point out that  the latter  (Feynman-Kac)  formula admits a  generalization to L\'{e}vy  processes, provided  we pass to the transition  kernel of the semigroup  $\exp (-t\hat{H})$ with    $\hat{H} = \hat {H}_{\alpha } =  |\Delta |^{\alpha /2} + V$. Then, the path measure needs to be  adopted to the jump-type setting, with sample paths of the  L\'{e}vy process replacing the Wiener ones,  see e.g. \cite{kaleta,kaleta1,lorinczi,lorinczi0}.  The  pertinent  expectation is taken with respect to  the path measure of the  $\alpha $-stable  process.

  We note that in Ref. \cite{kaleta} the function $V(x)$ is interpreted as delineating a potential landscape in which the random motion takes place.

\subsection{Conditioned  L\'{e}vy flights.}

We have not yet   specified any restrictions upon the properties  of the potential function $V(x)$, nor its  concrete functional form.  In the present paper, the potential is expected to be a  continuouus function and show up definite confining properties, which we adopt   after  \cite{kaleta}, by demanding  $\lim_{|x| \rightarrow \infty}  V (x) = \infty $.  Then, the Hamiltonian operator $\hat{H}_{\alpha }$ generically admits a   positive  ground state  function  $\varphi _0(x)$  with an isolated bottom eigenvalue  $\lambda _0$ (typically a  fully discrete spectrum is admitted).

With this proviso, let us invert our reasoning and consider evolution equations (24) and (25), with initial/terminal data $g(x)$ and $f(x)$ respectively.  This specifies the Schr\"{o}dinger interpolation of $\rho (x,t) = \theta (x,t)\theta _*(x,t)$ in the time interval $[0,T]$.

At this point, we shall narrow the generality of the  addressed  Schr\"{o}dinger boundary data and interpolation problem, by assuming that  actually for all  $t\in [0,T]$ we have
\be
\theta (x,t)= \exp (t \lambda _0)\,  \varphi _0(x)
\ee
where $\hat{H}_{\alpha } \varphi _0(x)  = \lambda _0 \varphi _0(x)  $.  Accordingly (we interchangeably use  $|\Delta |^{\alpha /2}$ or $(-\Delta )^{\alpha /2}$):
\be
 {\cal{V}}(x)=  V(x) - \lambda _0 = -   {\frac{1}{\varphi _0 (x)}} \,  (-\Delta )^{\alpha /2} \varphi _0 (x).
\ee

We point out a formal  appearance of the  potential function ${\cal{V}}(x)$ in the specific form, repeatedly invoked in our earlier papers,  ${\cal{V}}(x)=  - \rho _*^{-1/2} |\Delta |^{\alpha/2} \rho _*^{1/2}$, here specialized to the Cauchy case $\alpha =1$.

Let us indicate that the subtraction of $\lambda _0$ from the potential  $V(x)$ is a standard way to assign the eigenvalue zero to  the ground state function  $\varphi _0(x)$ of the "renormalized"  Hamiltonian $\hat{H} - \lambda _0$,   c.f.   \cite{kaleta,gar4,vilela,faris}.  In reverse,  it is the functional form of the right-hand-side of Eq. (27),  which guarantees that   $ \hat{H}_{\alpha }^{ren}=  \hat{H}_{\alpha } - \lambda _0=  (-\Delta )^{\alpha /2} +  {\cal{V}}(x)$  actually assigns the eigenvalue zero  to the eigenfunction  $\varphi _0(x)$.

Interestingly, a substitution of (26) to  Eq.  (19) implies
\be
p(y,s,x,t) = \exp[ \lambda _0 (t-s)]\,    k(y,s,x,t) {\frac{\varphi _0 (x)}{\varphi _0(y)}}.
\ee
This is a canonical functional form of the transition probability density  for so-called ground state  transformed jump-type  process,  whose probability density function  $\rho (x,t)$   asymptotically relaxes to
\be
\rho _*(x) =  {\frac{\varphi ^2_0 (x)}{ \int \varphi ^2_0(y) \, dy }}.
\ee

A detailed analysis of a number of exemplary cases can be found in Refs. \cite{gar3,gar4,gar5,gar6,vilela,faris}, where $\rho _*^{1/2}(x)$  notationally replaces $\varphi _0 (x)/\sqrt{\int \varphi ^2_0(y)dy}$. Accordingly, the compatibility condition   (27)  (the functional form of ${\cal{V}}(x)$ determines the functional form of $\rho _*^{1/2}(x)$ and in reverse)   reads:  ${\cal{V}} = -  (|\Delta |^{\alpha /2} \rho _*^{1/2}) / \rho _*^{1/2}  $.\\

{\bf Remark 2:} In the mathematical literature, \cite{kaleta,kaleta1,lorinczi},  the transition probability density (28) is  usually   transformed to the symmetric form (presuming  $k(t,x,y)=k(t,y,x)$):
\be
\tilde{p}(t,x,y) = \tilde{p}(t,y,x)=  \frac{e^{\lambda _0t}  k(t,x,y)}{ \varphi _0(x) \varphi _0(y)}
\ee
with the proviso that the appropriate function space is not $L^2(R, dx)$ but $L^2(R, \varphi _0^2(x) dx)$. Accordingly, $f(x,t) = E_x[X(t)] =  \int _R f(y) \tilde{p}(t,x,y) \varphi _0^2(y) dy =  \int _R f(y) p(t,x,y) dy$.
Here (take care of the interchange $x\leftrightarrow y$, since the symmetry is lost)
 \be
 p(t,x,y) = e^{\lambda _0 t}\,    k(t,y,x) {\frac{\varphi _0 (y)}{\varphi _0(x)}} \neq p(t,y,x),
 \ee
  c.f.  Eq. (28).

\subsection{Condition of detailed balance.}
Let us rewrite the defining formula (20) for $\rho (x,t)$  in the  familiar   form
 \be
 \rho (x,t)= \rho _*^{1/2} (x) \Psi (x,t),
 \ee
  where  $\rho _*(x)$ is defined according to (29), and
\be
\Psi (x,t) = \exp(\lambda _0 t)  \theta _*(x,t)   \sqrt{\int \varphi ^2_0(y)dy}.
 \ee
 In virtue of $\partial _t\theta _*= - \hat{H} \theta _* $, where $\hat{H} = \hat {H}_{\alpha } =  |\Delta |^{\alpha /2} + V$,   we  realize that
  $\partial _t \Psi = -(\hat{H}  - \lambda _0) \Psi $.

Consequently, the    associated fractional  Fokker-Planck equation, while adjusted to the present non-Langevin setting,   takes the form:
  \be
\partial _t \rho =   L^* \rho (x,t)=     \rho _*^{1/2}\, \partial _t \Psi  = [ - \rho _*^{1/2}(\hat{H}  - \lambda _0) \rho _*^{-1/2}]\, \rho (x,t)
\ee
in which we have encoded the  similarity  transformation,  \cite{kaleta,kaleta1,vilela,faris},   relating   the fractional  Fokker-Planck operator  $L^*$ and   $\hat{H}- \lambda _0$:
  \be
  L^* \equiv   - \rho _*^{1/2}(\hat{H}  - \lambda _0) \rho _*^{-1/2}.
  \ee

We   recall  that the time evolution of the  probability density function $\rho(x,t)$ is  governed by the  (adjoint) semigroup  $T^*_t = \exp (L^*t)$:  $T_t^* \rho (x,t) = \int p (t,y,x) \rho (y) dy$, where the transition pdf is given by the formula  (28).  Remembering that  Eq.(33) is an operator expression, where  the  action  of operators needs to be  read out from right to left, we can  extend  this identity  to the semigroup   operator  itself:  $T^* = \rho _*^{1/2}\, \exp (\hat{H}  - \lambda _0) \,  \rho _*^{-1/2}$.  We note that  $p(t,y,x)$    is an integral kernel of $T^*_t$, while   the entry    $\exp(t\lambda _0) \,  k(t,y,x)$  in the definition   (28)  of $p(t,y,x)$  actually is  an integral kernel of      $\exp [-(\hat{H}  - \lambda _0)t ]$.

 On the other hand  $T_t f(x)  = E_x[f(X_t)] =\int_R p(t,x,y) f(y)  = f(x,t)  $ is a  conditional expectation value of an "observable"  $f(x)$, evaluated  over   endpoints $X(t)=y$  of   sample paths  started  from   $x =X(0)$. Here $p(t,x,y)dy$   is interpreted as a probability of getting from $x=X(0)$ to the  vicinity of a point $y = X(t)$.    We point out that  $p(t,x,y)\neq  p(t,y,x)$, c.f.  (28), while generically $k(t,x,y) = k(t,y,x)$.

 Since $p(t,y,x)dx$  quantifies a probability of getting from $y$ to the  ($dx$)  vicinity of $x$ at time $t$, by employing (28), we can verify that in the present case the {\it  condition of  detailed balance}    manifestly holds true, c.f. \cite{gar1,balance}:
 \be
p(t,y,x) \rho _*(y) = p(t,x,y) \rho _*(x).
\ee
This, in conjunction with a redefinition of (28) according to $ p(t,y,x) \rightarrow p(t,x,y)$.  At this point we mention that for Langevin-driven L\'{e}vy processes the condition of detailed balance does not hold true, \cite{balance}.  Our non-Langevin approach has the detailed balance property built-in from the start, see e.g. \cite{gar1}  and references therein.

The      generator  $L$   of the pertinent jump-type process  appears in the form, c.f. also  \cite{kaleta,kaleta1},
   \be
    L\equiv  -  \rho _*^{-1/2}(\hat{H}  - \lambda _0) \rho _*^{1/2}.
    \ee
 which  conforms with  the   identity $L = \rho _*^{-1} L^* \rho _*$.

\subsection{Generators of conditioned L\'{e}vy flights.}

As yet, we have no detailed integral expressions for the motion generators  $L$ and $L^*$.
Let us begin from the   evaluation of the integral form for $L$, \cite{lorinczi1},
 which has been mentioned  elsewhere \cite{kaleta1}, but its derivation has been skipped.
To this end we shall resort to the regularized definition  (6) of the fractional Laplacian.
We are not aware of any simple computation method  starting from the  Cauchy principal value
 definitions (4) or (5), compare e.g. also \cite{cufaro}  and \cite{klauder}.

\subsubsection{Integral form of $L$, \cite{lorinczi1}.}

We shall follow  the notation of section II.B. Accordingly, for $\hat{H} = \hat {H}_{\alpha } =  |\Delta |^{\alpha /2} + V$, we have the eigenvalue equation  $\hat{H} \varphi _0(x)  = \lambda _0 \varphi _0(x)  $. Since $\lambda _0$ is interpreted as the bottom (ground state) eigenvalue, we readily infer Eq. (27), and thus
the action of $L$ upon any (suitable)   function  $f(x)$may be reduced to the evaluation of
\be
   \varphi _0 (L f) =   |\Delta |^{\alpha /2} (\varphi _0 f) - f (|\Delta |^{\alpha /2}\varphi _0)
\ee

The  action of the fractional Laplacian upon functions in its domain can be given in the integral form, and to this end we refer to  the regularized definition given in  Eq. (6). We have ($f(x)\equiv f(x,t)$ to be kept in mind):
\be
 (|\Delta |^{\alpha /2} f)(x) =  -  {\frac{1}2} \int_R [f(x+y) + f(x-y) - 2f(x)]\,  \nu (dy)
\ee
and therefore  (remember that we evaluate the above expression for $f(x)$ and subsequently for  the product
  $\varphi _0(x) f(x)$:
\be
 - \varphi _0(x)  (L f)(x)  =  {\frac{1}2} \int_R \{ \varphi _0(x+y)[f(x+y)- f(x)] + \varphi _0(x-y)[f(x-y) - f(x)]\} \nu (dy).
\ee

We change the variables under the integral sign to $z= x+y$ and  $z=x-y$ respectively, and next use the property $\nu(z-x)=\nu (x-z)$ of the L\'{e}vy-stable measure $\nu(dy)=\nu(y) dy$. The outcome is:
\be
 - \varphi _0(x)  (L f)(x) = {\frac{1}2}  \int_R [f(z) - f(x)] \varphi _0(z) \nu (z-x) dz
\ee
i.e. the integral form of $L$ reads
\be
(Lf)(x) = \mathcal{A}_{\alpha} \int _R {\frac{f(x) - f(z)}{|z-x|^{1+\alpha }}}  {\frac{\varphi _0(z)}{\varphi _0(x)}} dz,
\ee
to be compared with the  standard  fractional Laplacian   definition  (5).

\subsubsection{Integral form of $L^*$.}

Let us rewrite the transport equation (34)  in the notation compatible with that used in the previous subsection. We have:
\be
\partial _t \rho =   L^* \rho =     - \varphi _0 (\hat{H}  - \lambda _0) {\frac{1} \varphi _0}\, \rho .
\ee
By employing the eigenvalue equation (27) and (34),  we arrive at  (here $\rho (x) \equiv \rho (x,t)$):
\be
L^* \rho = - \varphi _0 (|\Delta |^{\alpha /2} {\frac{\rho }{\varphi _0}})  +  {\frac{\rho }{\varphi _0}}  (|\Delta |^{\alpha /2} \varphi _0)
\ee
to be compared with (38).  Basically we can repeat major  steps of the previous  evaluation.

By employing (39), properly  rearranging terms  and executing suitable changes of integration variables, we get:
\be
L^* \rho (x)=\int _R \left[ {\frac{\varphi _0(x)}{\varphi _0(z)}} \rho (z) - {\frac{\varphi _0(z)}{\varphi _0(x)}} \rho (x)\right] \nu(z-x) dz,
\ee
where $\nu (z-x)  = {\cal{A}}_{\alpha } /|z-x|^{1+\alpha }$.   It is instructive to compare this result with an alternative derivation, based on the definition (3)  of the fractional Laplacian, c.f. Eqs. (83), (84) in Ref. \cite{klauder}.

 {\bf Remark 3:}  The same formula can be obtained by invoking a direct construction  of L\'{e}vy processes whose confinement is due to the response to  potentials rather than to  conservative  forces   proper, \cite{sokolov}, see also \cite{gar,gar1,gar2}.  Indeed, the relevant formula (28) in Ref. \cite{sokolov} has the form:
  \be
\partial _t \rho (x,t) = {\cal{A}}_{\alpha } \{ s(x)  \int  {\frac{q(y,t) - q(x,t)}{|x-y|^{1+\alpha }}} dy
  - q(x,t) \int  {\frac{s(y) - s(x)}{|x-y|^{1+\alpha }}} dy \}
\ee
where $s(x) = e^{-\Phi (x)/2} \equiv  \varphi _0(x)$, while $q(x,t) = \rho (x,t)/s(x)$, and upon suitable rearrangements is identical with Eq. (45).\\
It is the   salience  field $s(x)$,   or  (in view of associations with the notion of the  Boltzmann equilibrium pdf)
the  (would-be Boltzmann)  potential function $\Phi (x)= -2 \ln s(x)$, which receives an interpretation of the
salience or potential landscape respectively  in Ref. \cite{sokolov}, see also \cite{gar,gar1,gar2}.  An alternative potential/energy  landscape notion is associated with the related Feynman-Kac potential, \cite{kaleta}.

\section{Cauchy process in the interval: Superharmonic  approximation of Dirichlet boundaries.}

\subsection{Reference  spectral  data  of the Cauchy    generator   $|\Delta |^{1/2}_{\cal{D}}$   in the
 infinitely deep potential well  (interval).}

The  main problem,  which  hampers a  usage of  L\'{e}vy flights  as  computationally  useful model systems,   is the nonlocality of the stochastic  process itself and the nonlocality of its  generators.  Analytically tractable  reasoning  is   seldom  in the reach and  one needs to rely on  computer-assisted methods.

For L\'{e}vy flights  in  the interval  with absorbing  endpoints  (exterior Dirichlet boundary conditions are necessary here),  approximate analytic formulas  are  available  for the  spectral relaxation    data (fractional Laplacian   eigenvalues and  eigefunctions).  The approximation  accuracy has been significantly improved  by resorting to numerics, specifically   in the Cauchy case,   \cite{kwasnicki1,malecki,zaba,zaba1,zaba2,zaba3}. The  lowest eigenvalues and eigenfunctions shapes of the Cauchy Laplacian  in the interval  (with Dirichlet boundaries)   were obtained by  different computer-assisted   methods,  with a high degree of  congruence, c.f. comparison Tables in Ref. \cite{zaba1} and references therein.

 Let $D$ be an open set in $R$, like e.g. the open interval$(-1,1)$.   The  Dirichlet boundary  condition actually takes the form of the exterior restriction, imposed upon  functions  in the domain of the fractional Laplacian:
\be
 |\Delta |^{\alpha /2}  \psi _n(x)= \lambda_n \psi _n(x),
 \ee
 for all $x\in D$ while  $\psi _n(x)=0$ for all $x\in R\setminus D$ and $\psi \in L^2(D)$.
    We deal here with the exterior Dirichlet condition valid on the complement of $D$ in $R$. This  should be
  contrasted with the standard Brownian case, where the Dirichlet condition is imposed locally at the  boundary  $\partial D$ of $D$, so that $D\cup \partial D = D^c$ is a closed set (interval with endpoints, like $[-1,1]$).

  Here   $\lambda _n >0$  for all  natural  $n\geq  1$, \cite{kulczycki}-\cite{zaba3}.  The eigenfunctions $\psi _n(x)$ are continuous and bounded in $D$,  and reach the boundary  $\partial D $ of $D$ continuously, while approaching the (Dirichlet) boundary  value zero. The ground state function $\psi _1(x)$ is strictly positive in $D$.
    The fractional Laplacian  with Dirichlet boundary conditions we name  the  Dirichlet fractional Laplacian, and abbreviate to the notation $|\Delta |^{\alpha /2}_{\cal{D}}$.

  There exists  an analytic estimate for the spectrum     of   $|\Delta |^{\alpha /2}_{\cal{D}}$ in case  of arbitrary stability index $0< \alpha <2$, \cite{kwasnicki1}. For all $n\geq 1$ we have
   \be
   |\lambda _n - [{\frac{n\pi }2} - {\frac{(2- \alpha )\pi }8}]^{\alpha }| \leq {\frac{2- \alpha }{n\sqrt{\alpha }}},
  \ee
   but the approximation accuracy may be considered reliable beginning roughly  from   $n\geq 10$, c.f.  \cite{zaba2,zaba3}.    For reference purposes, we indicate that the lowest two eigenvalues  read   $\lambda _1= 1.157791$ and $\lambda _2= 2.754795$, which should be set against two  bottom eigenvalues of the standard Dirichlet Laplacian $(- \Delta )_{\cal{D}}$ equal $\pi ^2/4= 2.4674$ and $\pi ^2= 9.8696$ respectively, \cite{trap,zaba3}.  In the Cauchy case, the spectral gap  $\lambda _2-\lambda _1$   is much lower than this in the Brownian case, c.f. \cite{kulczycki}.

We have  found    quite accurate analytic   approximation formulas for the lowest eigenfunction shapes, valid  for any $0<\alpha <2$, c.f. \cite{zaba1,trap}:
 \be
\psi _1(x)= C_{\alpha , \gamma} [(1-x^2) cos(\gamma x)]^{\alpha /2},
\ee
where $C_{\alpha , \gamma}$ stands for the $L^2(D=[-1.1])$ normalization factor, while $\gamma $ is considered to be the "best fit" parameter.   This  analytic  formula for  the ground state function, well conforms with numerically simulated curves, \cite{zaba,zaba1,zaba2,zaba3,trap}.

 In the Cauchy case, $\alpha =1$, almost prefect fit (up to the available graphical resolution limit)
  has been obtained for  $\gamma = {\frac{1443}{4096}}\pi $, with $C= 0.92175$, \cite{zaba1}.
   It is known that all eigenfunctions show the $\sim \sqrt{(1-x^2)}$ decay to $0$, while approaching
   the interval $[-1,1]$ endpoints, see e.g. \cite{kulczycki,zoia,zaba,zaba1}.

Technical details are available in Refs.  \cite{zaba,zaba1}, where  we have devised  the method of solution of the Schr\"{o}dinger-type spectral   problems  by means of the Strang  splitting technique  for  semigroup operators. The method has been   comparatively  tested   by  referring to the analytically solvable   Cauchy oscillator  model, and  next employed  in the Cauchy well  setting to  deduce the lowest eigenvalues and  eigenfunctions shapes   of the Cauchy - Dirichlet Laplacian on the interval.  The analysis has been complemented by executing the sequential approximation of the Cauchy infinite well in terms of the  deepening  finite well problems.

 We note that in contrast to the locally defined boundary data in the Brownian case, the Cauchy operator  (and likewise other $\alpha $-stable generators), in view of its nonlocality, needs an exterior  Dirichlet condition. Accordingly, functions from the operator domain need to vanish on the whole complement $R\setminus  D $  of the open set $D=(-1,1)$  (the  closure of  $D$  is  $D^c=[-1,1]$).

\subsection{Non-Langevin approach.}

Since, the Cauchy generator  in the interval with absorption at the endpoints,  is  spectrally identical with that of the infinite  (quantum association) Cauchy well i.e. the Cauchy operator    with exterior Dirichlet boundary data  $|\Delta |^{1/2}_{\cal{D}}$  (${\cal{D}} \subset R$),  \cite{kulczycki,zaba,zaba1},  it is natural to address an issue of   its  sequential approximation    by     superharmonic Cauchy-Schr\"{o}dinger  operators    $\hat{H}= |\Delta |^{1/2} + V(x)$ with $V(x)= x^{2m}$, $m \geq 1, m \to \infty $,   defined on $R$.

This sequential procedure stems from the large $m$ properties of the potential function $V(x)\sim x^m$, with $m$ even.  One tacitly presumes that in the $m\to \infty $ limit, defining properties of the infinite well
enclosure set on $[-1,1]\subset R$ are reproduced.

 We point out some obstacles, that need to be carefully handled   in connection with   the point-wise  $m \to \infty $ limit. Namely, (i)   $x^m$, at $x=\pm 1$   takes the value 1 for all $m$ ,
(ii) $x^m/m$ takes the value  $1/m \to  0$  as $m \to \infty $, while  (iii) $m x^m$ takes the value
$m \to \infty $.   In all three cases, the point-wise limit $m\to \infty $ establishes the walue $V(x)\to 0$
for  $|x|<1$ and $V(x)\to \infty $ for $|x|>1$.  We emphasize the  relevance of the  exterior (to the interval $(-1,1)$) property of $V(x)$,  it diverges to infinity everywhere on $R\setminus (-1,1)$.

At this point it is useful to mention the traditional definition of the infinite well enclosure, which is is considered in conjunction  with the concept of the Dirichlet boundary data: $V(x) = 0, |x|<1$ and $V(x) = \infty , |x|\geq 1$. Evidently, it is  $V(x) = mx^m$ which consistently approximates this enclosure for $m\to \infty$, c.f. also Ref. \cite{brownian}.

  We point out that  sequential {\it  finite}  well approximations  were studied in detail in Refs. \cite{zaba,zaba1}.  A complementary  analysis of  finite well   approximations of the standard Laplacian (and  the confined Brownian motion)   in the interval can be found in Ref. \cite{brownian}. See also \cite{trap} for a general   discussion    of L\'{e}vy flights in bounded domains.  In all these case the boundary value of the limiting infinite well potential has been  assumed to be   equal  $\infty $.

To check the validity of the sequential superharmonic approximation of   $|\Delta |^{1/2}_{\cal{D}}$,    we have generalised the method of Ref. \cite{zaba}, originally   employed  to obtain the  spectral solution of the  Cauchy oscillator.  The superharmonic   semigroup   generator reads $\hat{H}=  |\Delta |^{1/2} + x^{2m} $. In view of the  implicit "ground state reconstruction strategy", we are interested in the  solution of   $\hat{H} \psi _1  = \lambda _1 \psi _1$, with  $\psi _1$ given in the $L^2(R)$ normalized form   $\psi _1 = \varphi _0 (x)/\sqrt{\int \varphi ^2_0(y)dy}$  of Section II.C, where $\lambda _1$ is   (in the present case) the  positive   bottom eigenvalue.

Computer assisted outcomes are displayed  in Fig.1  and show definite convergence symptoms  towards spectral data of the infinite Cauchy well.   We depict   a  sample of  $L^2(R)$-normalized ground state  functions of $\hat{H}= |\Delta |^{1/2} + x^{2m}$  for  $m=1,4,10, 50, 250, 2500$.   The $m=2500$  curve  (yellow) is indistinguishable, in the adopted resolution scale,  from the best-fit  infinite Cauchy well  eigenfunction  (black)   $\psi _1(x)= 0.921749 [(1-x^2) \cos(1443\pi x/4096)]^{1/2}$ of Ref. \cite{zaba1}, see also Eq. (47).

 The bottom eigenvalue dependence on  the superharmonic exponent $2m$ is  depicted  as well. Convergence symptoms towards the infinite Cauchy  well eigenvalue $E_1= 1.157791$ (dashed line level) are conspicuous. The last displayed     eigenvalue  (circle) corresponds to $2n= 5000$ and reads  $E_1(5000) \sim 1.55232$.\\

{\bf Remark 4:}  Since for all $m\geq 1$, the spectrum of $\hat{H}= |\Delta |^{1/2} + x^{2m}$ is positive, with a bottom eigenvalue $\lambda _1[m]$, it is clear that $\hat{H} - \lambda _1[m]=  |\Delta |^{1/2} + {\cal{V}}(x)$, where ${\cal{V}}(x) = -  [|\Delta |^{1/2}  \psi _1 (x)] / \psi _1(x) $, assigns the bottom eigenvalue zero  to the  positive  eigenfunction $\psi _1(x)$.  We point out a notational change:  $\lambda _0$ and $\psi _0(x)$ of Section II are now replaced by $\lambda _1$ and $\psi _1(x)$ respectively.

\begin{figure}
\begin{center}
\centering
\includegraphics[width=80mm,height=80mm]{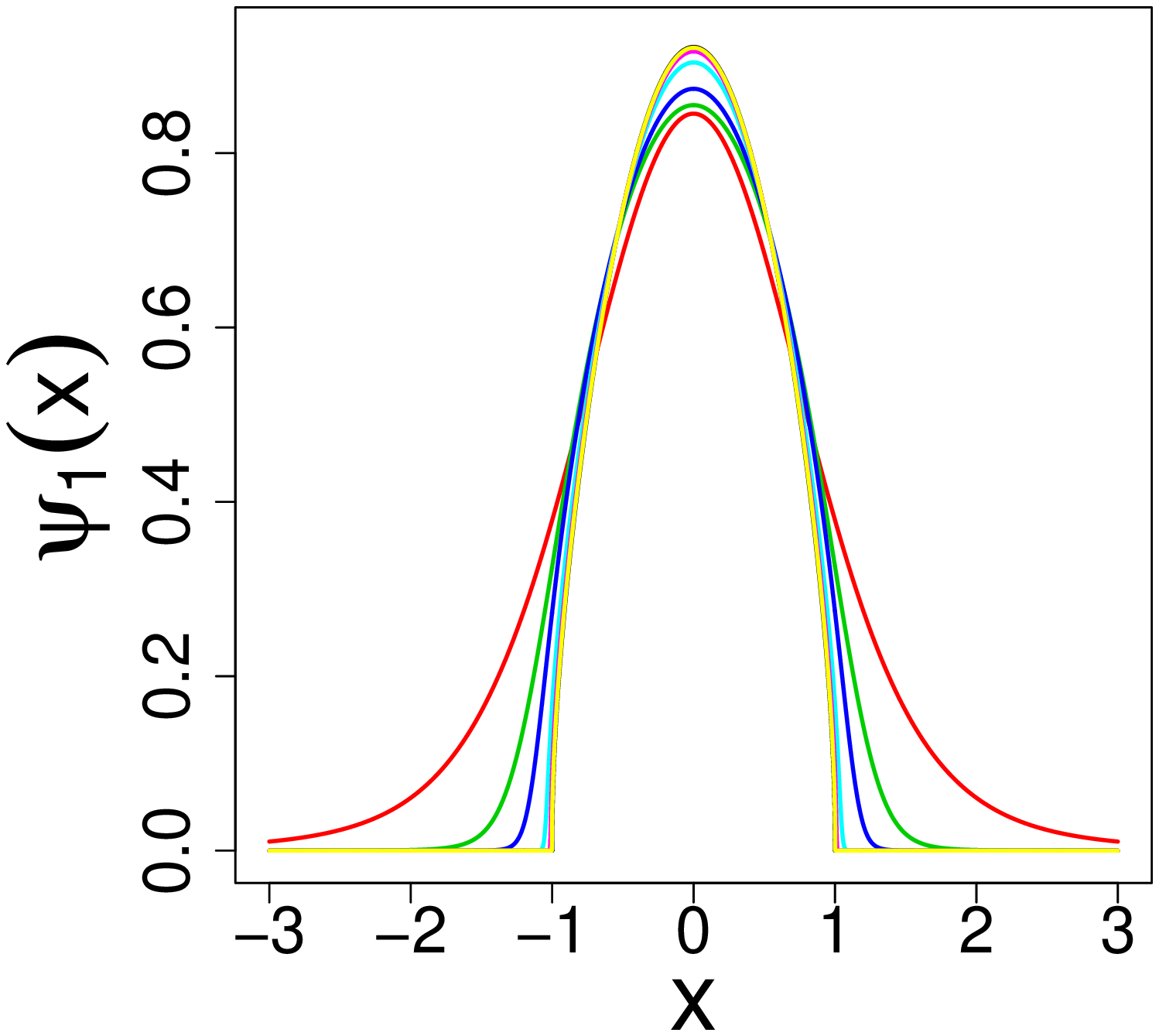}
\includegraphics[width=80mm,height=80mm]{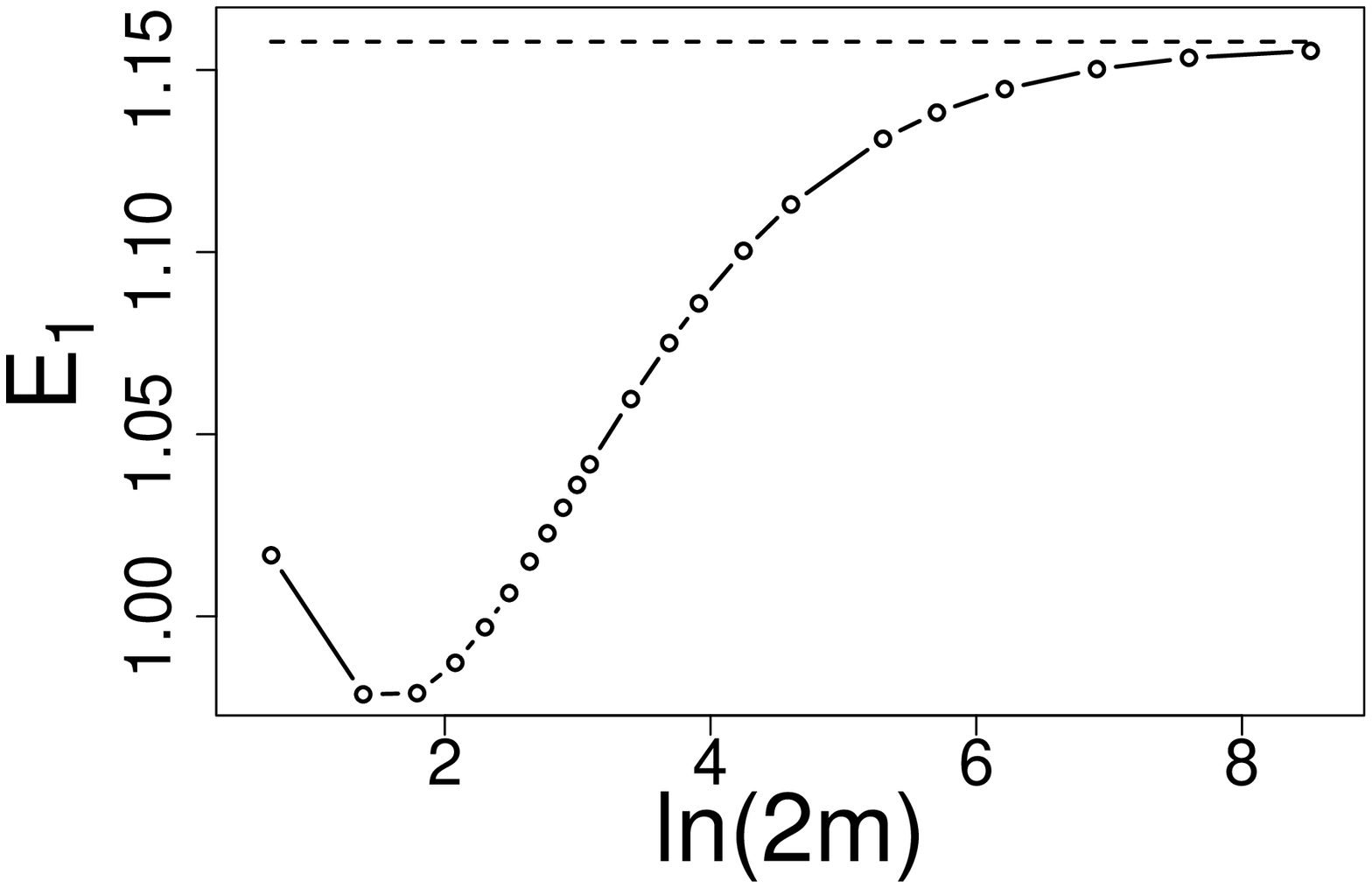}
\caption{Left panel:  $L^2(R)$-normalized ground state  functions of $\hat{H}= |\Delta |^{1/2} + x^{2m}$  for  $m=10$ (red), $50$ (green), $250$ (blue), $2500$ (yellow) and  the best-fit  infinite Cauchy well  eigenfunction  (black/yellow)  of Ref. \cite{zaba1}, c.f. Eq. (47) with $\alpha =1$.  Right panel: The bottom eigenvalue dependence on $2m$. The dashed line  sets  the energy  level    corresponding   to the infinite Cauchy  well eigenvalue $E_1 \sim 1.1578$.}
\end{center}
\end{figure}

\subsection{Reverse engineering: Langevin alternative.}

For each $m\geq 1$ let there be given $\rho _*(x)=  \psi _1^2(x)$,  where $\psi _1(x)$ is the  $L^2(R)$-normalized
positive-definite  ground state function of   the superharmonic Hamiltonian $\hat{H} = |\Delta |^{1/2} + x^{2m}$.
 As we know,  $\psi _1(x) = \rho ^{1/2}_*(x)$  determines  the   stationary  probability density  $\rho _*(x) = \psi ^2_1(x)$ of a Markovian stochastic process, obeying the principle of  detailed balance.  The  pertinent    random  dynamics can  be  recovered by following the non-Langevin approach of  Section II.

On the other hand,   given the very   same  stationary pdf  $\rho _*(x)$,   we may  attempt a reconstruction of the Langevin system, subject to the Cauchy noise,   that would yield a  relaxation  of any suitable $\rho (x,t)$  to  $\rho _*(x)$  (here considered as a   pre-specified "target", \cite{eliazar}).     We point out that  for Langevin-driven L\'{e}vy processes the condition of detailed balance  generically  does not hold true, \cite{balance,balance1}, while being  valid in the non-Langevin case.

To quantify   relaxation properties of  a Markovian L\'{e}vy-Langevin process, we  need to resort to Eqs. (6) and (7).   Actually,  to  recover  the appropriate  fractional   Fokker-Planck evolution  $\partial _t\rho = -\nabla (b\cdot \rho ) -  |\Delta |^{1/2}\rho  $   of  any suitable  $\rho (x,t)$,   we must reconstruct the functional form of the drift function $b(x)$.    Its functional form must be  compatible with  the assumption that   the  chosen target   $\rho _*(x)$  is a stationary solution  of  the Cauchy F-P equation.

 This amounts to evaluating (in Ref. \cite{eliazar}  an alternative   drift reconstruction  procedure has been  proposed, realized   on the level of   Fourier transforms)
\be
b(x)= -    {\frac{\int  |\Delta |^{1/2}\rho _*(x)\, dx}{\rho _*(x)}} \, .
\ee
The indefinite integral is to be numerically handled, since  we know    a priori the functional shape of $\rho _*(x)$  (numerical data are in hands). See e.g.  \cite{gar0}  for a couple of analytically accessible examples.

The integration  procedure is straightforward.  Given $\psi _1(x)$, we evaluate  point-wise the target pdf $\psi _1^2(x)= \rho _*(x)$. Next we evaluate numerically (c.f. \cite{zaba,zaba1} $|\Delta |^{1/2} \rho _*(x)$.   Since we know that with the growth of $m$,    $\rho _*(x)$   decays rapidly beyond the interval $[-1,1]$  (c.f. Fig. 1),
and likewise $|\Delta |^{1/2} \rho _*(x)$, the indefinite integral  of the form $\int f(t)dt$ is actually computed as a definite one  $\int_a^x (f(t)dt$, where  the  finite lower integration bound    $a< -1$  replaces  the "normal"  $-\infty $ in the integral.

 The outcome of computations if displayed in Fig.2, where all  depicted figures derive from the ground state function $\psi _1(x)$  of the Cauchy  operator $\hat{H}= |\Delta|^{1/2} +  x^{2m}$, where  $m>1$.  For comparison,  we depict  the drift for the Brownian motion  in the interval  (equivalently - infinite well)  with inaccessible boundaries,   \cite{gar4,brownian}, where $ b(x)=-\pi \tan(\pi x/2)$ and  $\psi _1(x)=\cos(\pi x/2)$ on $[-1,1]$.  The corresponding random motion belongs to the category of taboo processes, see  \cite{gar4,gar5,brownian}.  In Ref.\cite{brownian} a  comparative discussion can be found, of  affinities and  the incongruence  of    Brownian motion scenarios in the infinite well  enclosures, in case of (i)  absorbing boundaries  (Dirichlet), (ii) boundaries  inaccessible from the interior   (taboo version of  Dirichlet data),   or   (iii) impenetrable,  internally   reflecting (Neumann).

\begin{figure}[h]
\begin{center}
\centering
\includegraphics[width=85mm,height=75mm]{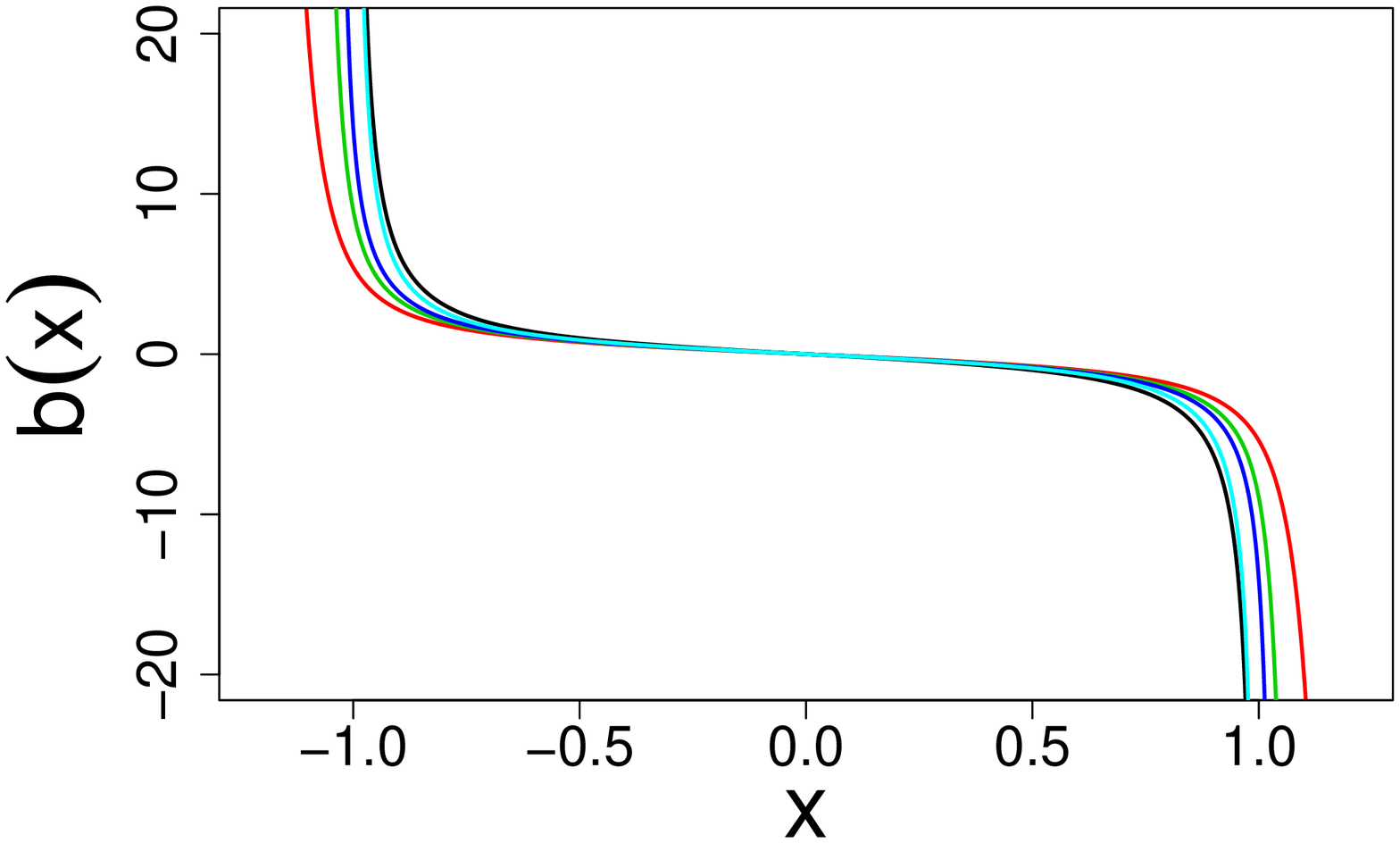}
\includegraphics[width=75mm,height=75mm]{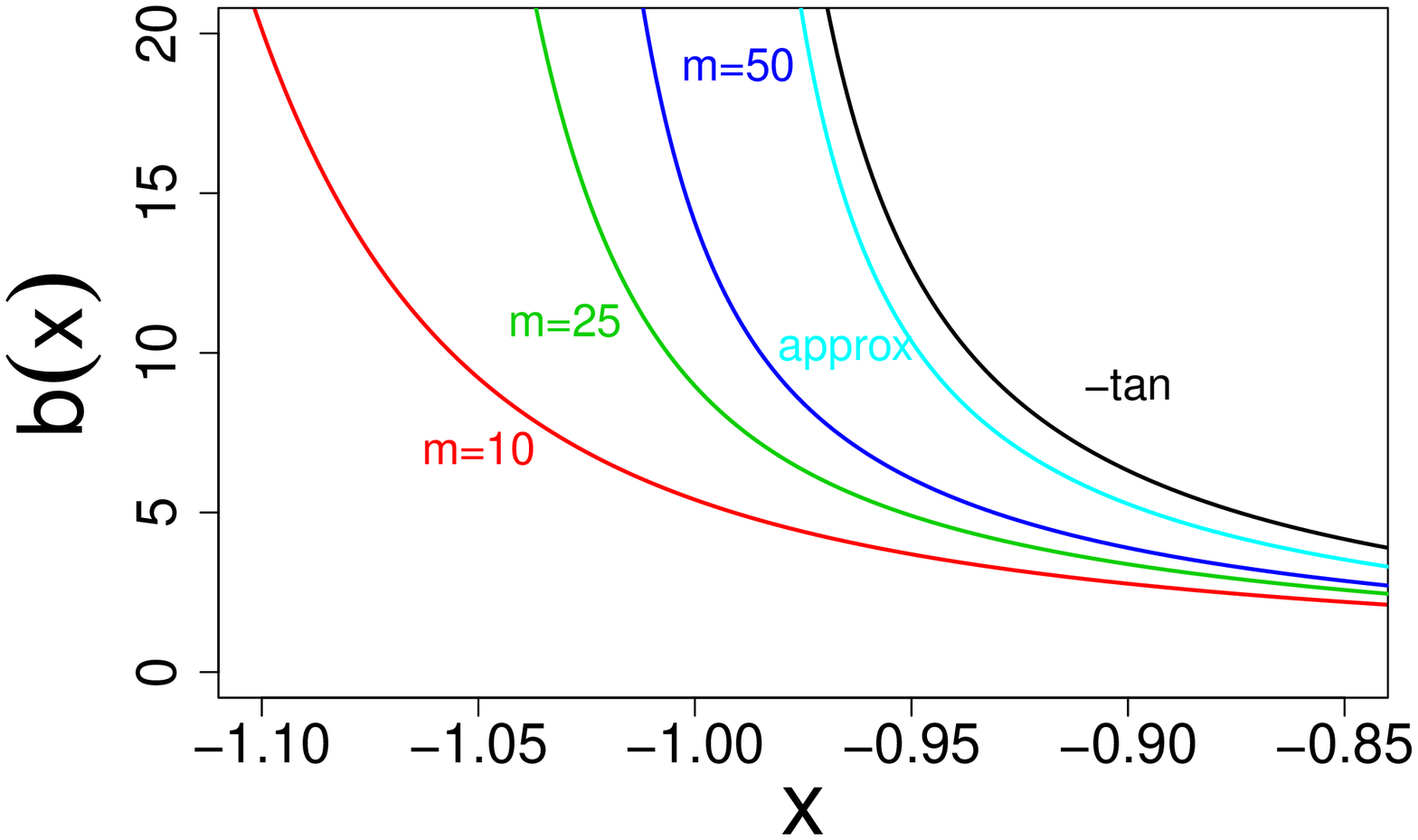}
\caption{Forward drift of the fractional  (Cauchy) Fokker-Planck evolution.   Left panel: a couple of  $b(x)$ curves is depicted. Details are displayed in  the right panel: $m=10$ (red), $m=25$ (green),  $m=50$ (blue), $approx$ (turquoise)  derives from the best-fit  approximate formula for $\psi _1(x)$ in the infinite well, Eq. (47) for $\alpha =1$, while $ - tan $ (black) refers to the  forward drift of the  Brownian motion  in the infinite well (interval) with  inaccessible boundaries (named  Brownian   taboo process), \cite{gar4}.}
\end{center}
\end{figure}

\section{Superharmonic  Cauchy-Langevin systems, their non-Langevin  partners  and "impenetrable   boundaries".}

\subsection{Superharmonic approximation of the Cauchy process in the interval (Langevin realization).}

While departing from the Langevin picture of Cauchy flights, which  are  confined by superharmonic potentials $U(x)= x^{2m}/2m, m\geq  1$   one arrives at the  $2m$  sequence of  fractional Fokker-Planck equations, whose stationary solutions  can be obtained in a closed analytic form.   To stay in conformity with  Refs. \cite{dubkov,dubkov1}   we use the notation $U(x)$ instead of $V(x)$,   in the  specific  context of Langevin-F-P drift fields  $ b(x)  \sim - \nabla U$. The notation $V(x)$,  and likewise  ${\cal{V}}(x)$,  is reserved exclusively  for Feynman-Kac potentials and  delineated by them   "potential landscapes".

 The  formal limiting   behavior   of the pertinent $2m$-sequence, as $m\rightarrow \infty $,  appears to be    interpreted     in   conjunction with the concept  of {\it reflected} L\'{e}vy flights  in the  interval, \cite{denisov,dubkov,dubkov1}.  This interpretation basically stems from a  suggestive  large $m$  behavior of  a superharmonic sequence of standard Fokker-Plack equations (for superharmonic, drifted Brownian motion), c.f. \cite{brownian,dubkov,dubkov1} and a semiclassical view of what a dynamics in the infnite well should  possibly  look like, with a traditional picture  of a reflecting ball moving with a uniform velocity between the impacts at interval endpoints (reflecting walls).

 The  reasoning  of Refs. \cite{dubkov,dubkov1} employs  the  Langevin-type  equation   $\dot{x}= b(x)  + B^{\alpha  }(t)$    with  a deterministic term $b(x)= - \nabla U(x) =  - x^{2m -1}, m\geq 1$   and   the  additive L\'{e}vy "white noise" term.   This  leads to   a fractional Fokker-Planck equation (\cite{fogedby},  governing the time evolution  of the pdf  $\rho (x,t)$  of the   relaxation process:  $ \partial _t\rho = - \nabla  (b \cdot \rho ) -  |\Delta |^{\alpha /2}\rho  $.

In Refs. \cite{dubkov,dubkov1}, for a particular choice of the Cauchy noise ($\alpha =1$),  an explicit form of the  stationary solution  has been derived for  all values of $m>2$. A formal $m \to \infty $ limit allows to reproduce  the Cauchy  version of the steady-state  solution for "L\'{e}vy flights in a confined domain",  \cite{denisov}, where actually it is considered as  "the case of stationary L\'{e}vy flights  in an infinitely deep potential well".  Since it is claimed by the Authors, that under the infnite well "confined geometry" conditions, "the origin of the preferred concentration of flying objects nears the boundaries in nonequilibrium systems is clarified", we point out our observations to the contrary, c.f. Section III and \cite{kwasnicki,zaba,zaba1,brownian}.

{\bf Remark 5:}  We point out that in the original notation of Ref. \cite{dubkov,dubkov1},  it is   $p_{st}(x)$  which  stands for our $\rho _*(x)$.  To simplify calculations, we scale away a parameter $\beta $ in the formulas (21), (22) of Ref. \cite{dubkov1} (this amounts  to  setting  $\beta =1$).  In the original formulas of Ref. \cite{dubkov1},   the  pertinent parametr $\beta$  comprises $m$ and the  interval  half-length $L$  in the proportionality factor:  $\beta \sim L^{2m/2m-1}$.  As $m \to \infty $, we have $\beta \to L$.  In the present paper we set  $L=1$   and   so preselect the interval $[-1,1]$ as a support for the limiting distribution.\\

For  odd values of $m=2k+1$  we have  \cite{dubkov1} the following expression for the stationary solution $\rho _*(x)$ of the   superharmonic fractional Fokker-Planck equation:
\be
\rho _*(x) =  {\frac{1}{ \pi (x^2+1)}}\prod _{l=0}^{k-1} {\frac{1}{x^4 - 2x^2 cos[\pi (4l+1)/(4k+1)] + 1}}
\ee
while for even values of $m=2k$, we have:
\be
\rho _*(x) = {\frac{1}\pi }  \prod _{l=0}^{k-1} {\frac{1}{x^4 - 2x^2 cos[\pi (4l+1)/(4k - 1)] + 1}}.
\ee
A representative  sample of pdf shapes  $\rho _*(x)$   for low values of $m\geq 5$ is depicted in Fig. 3, while a sample   of square root pdfs $\rho _*^{1/2}(x)$ for  larger (medium-sized)  superharmonic exponents ($m=10$ to  $m=100$) is displayed in Fig. 4.

\begin{figure}[ht]
\begin{center}
\centering
\includegraphics[width=75mm,height=75mm]{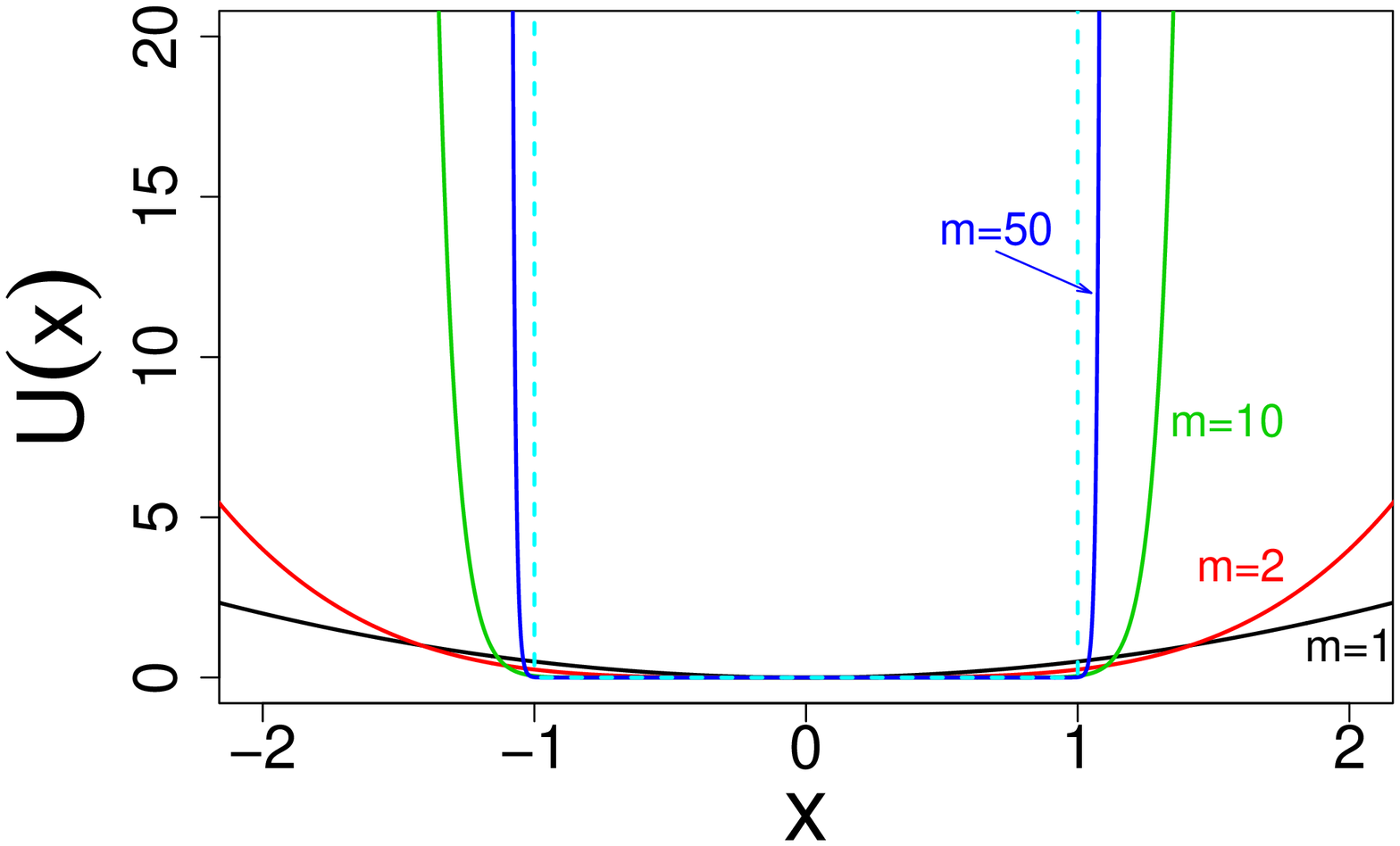}
\includegraphics[width=90mm,height=75mm]{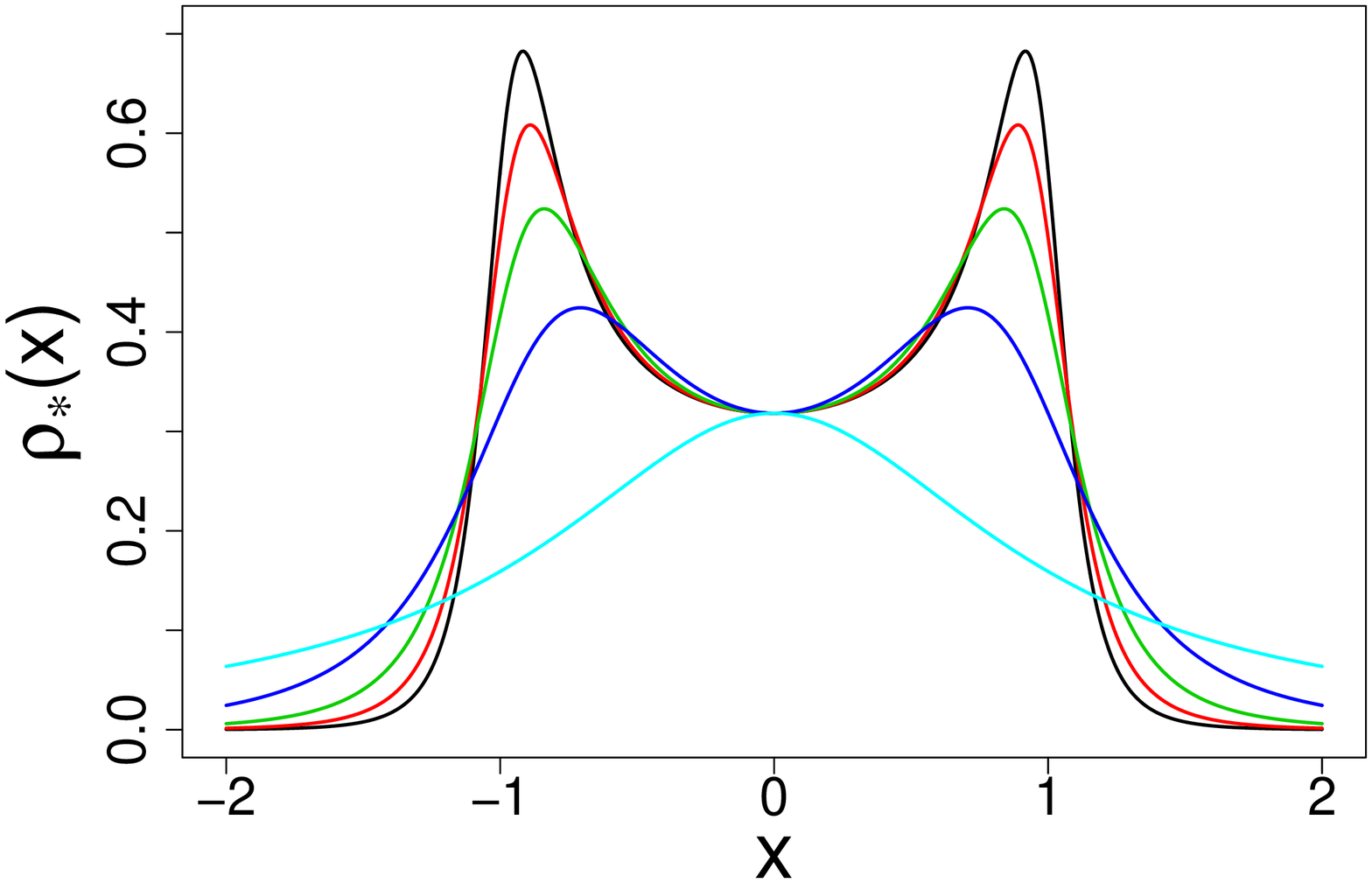}
\caption{Left panel: The superharmonic potential  $U(x)= x^{2m}/2m$ is depicted for $m= 1, 2, 10, 50$.
The dotted line indicates the shape of the infinitely deep well potential, supported
 on the interval $[-1,1]$ with the bottom energy level equal zero. We note that $U(\pm 1)=1/2m$
 goes to zero as $m \to \infty $.  Right panel: We depict stationary pdfs $\rho _*(x)$
  of the (Langevin-induced) superharmonic Cauchy - Fokker-Planck equation, for $m=1$ (turquois), $2$  (blue),
   $3$ (green), $4$ (red), $5$ (black).}
\end{center}
\end{figure}

\begin{figure}[hb]
\begin{center}
\centering
\includegraphics[width=75mm,height=75mm]{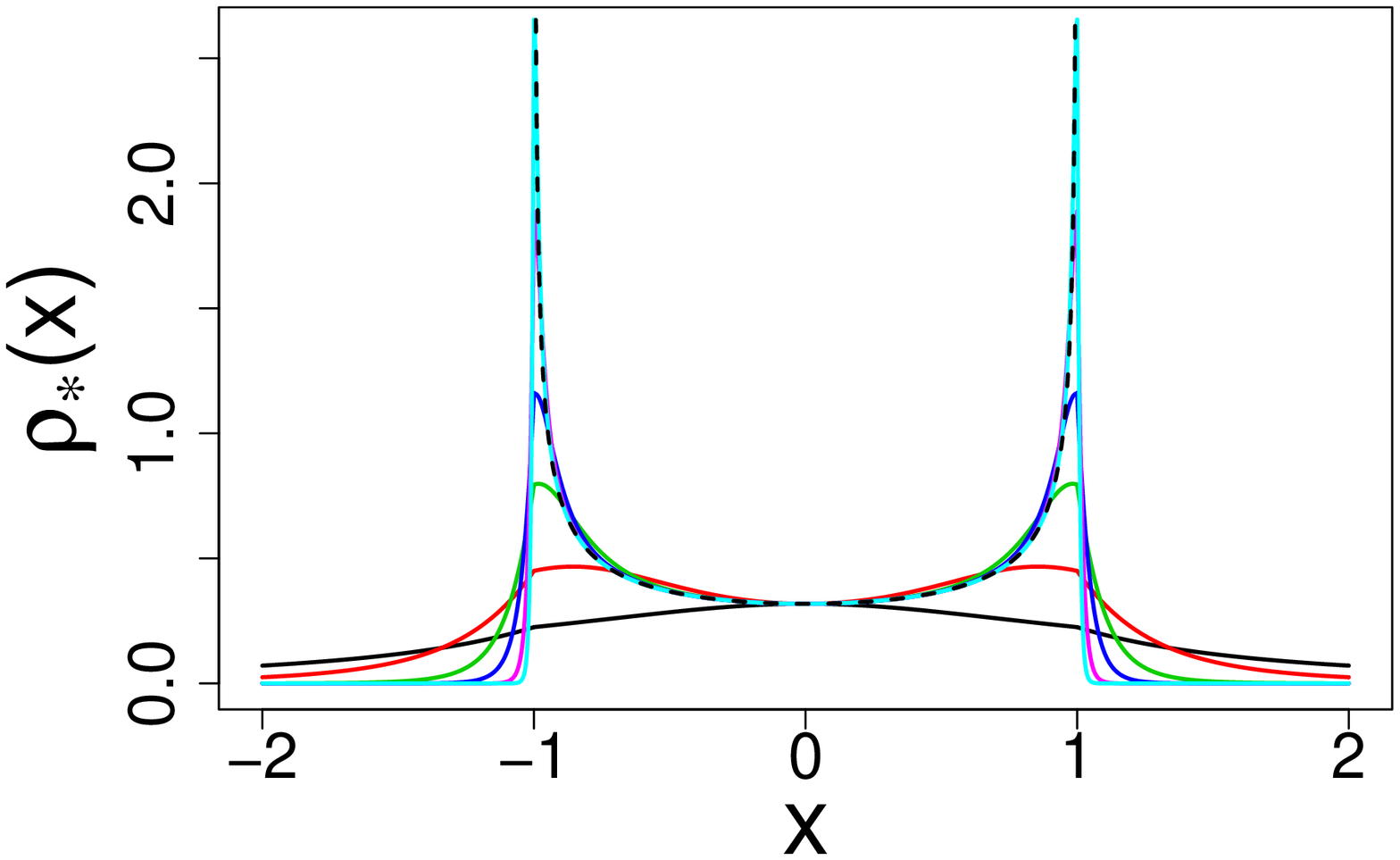}
\includegraphics[width=90mm,height=75mm]{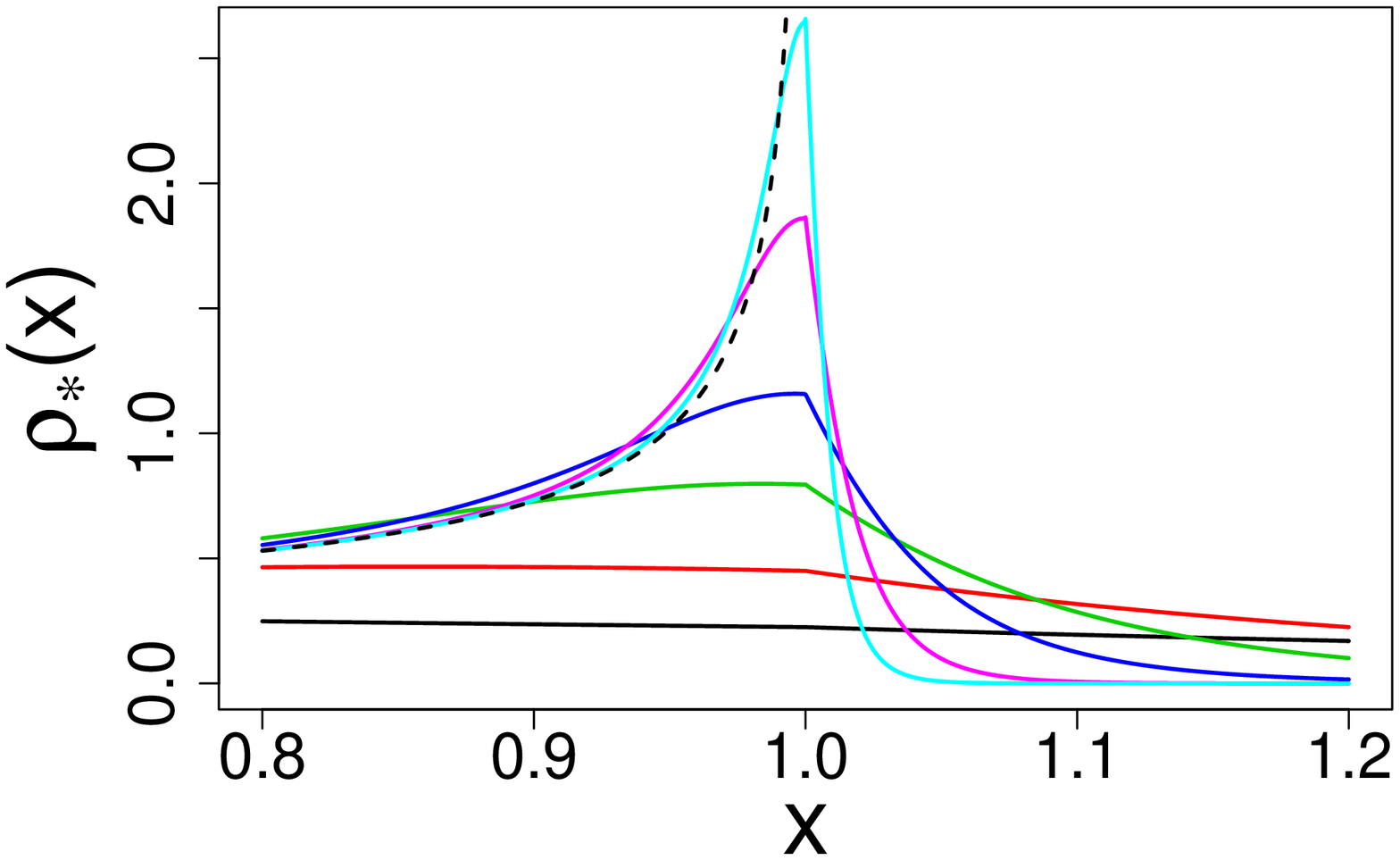}
\caption{Left panel: Superharmonic stationary probability densities $\rho _*(x)$ are depicted for  $m=1, 2, 5, 10,  25, 50$ (indices correspond to consecutive maxima in the  growing sequence).
The dotted line indicates the arcsine pdf $\rho _*(x)[m\to \infty ] =  1/ \pi \sqrt{1-x^2}$, which "lives" exclusively within the open interval $(-1,1)$ and is undefined   (actually does not exist) at $\pm 1$ .
 Right panel: enlarged vicinity of the endpoint $x=1$. Locations of   maxima (for each m-th pdf)
  relative to the   arcsine curve, are  clearly displayed: they are bounded from above by this curve. The latter is not en envelope, since for  all  finite  $m>1$ we encounter  two intersection points at every  arcsine branch.}
\end{center}
\end{figure}

\begin{figure}[ht]
\begin{center}
\centering
\includegraphics[width=85mm,height=80mm]{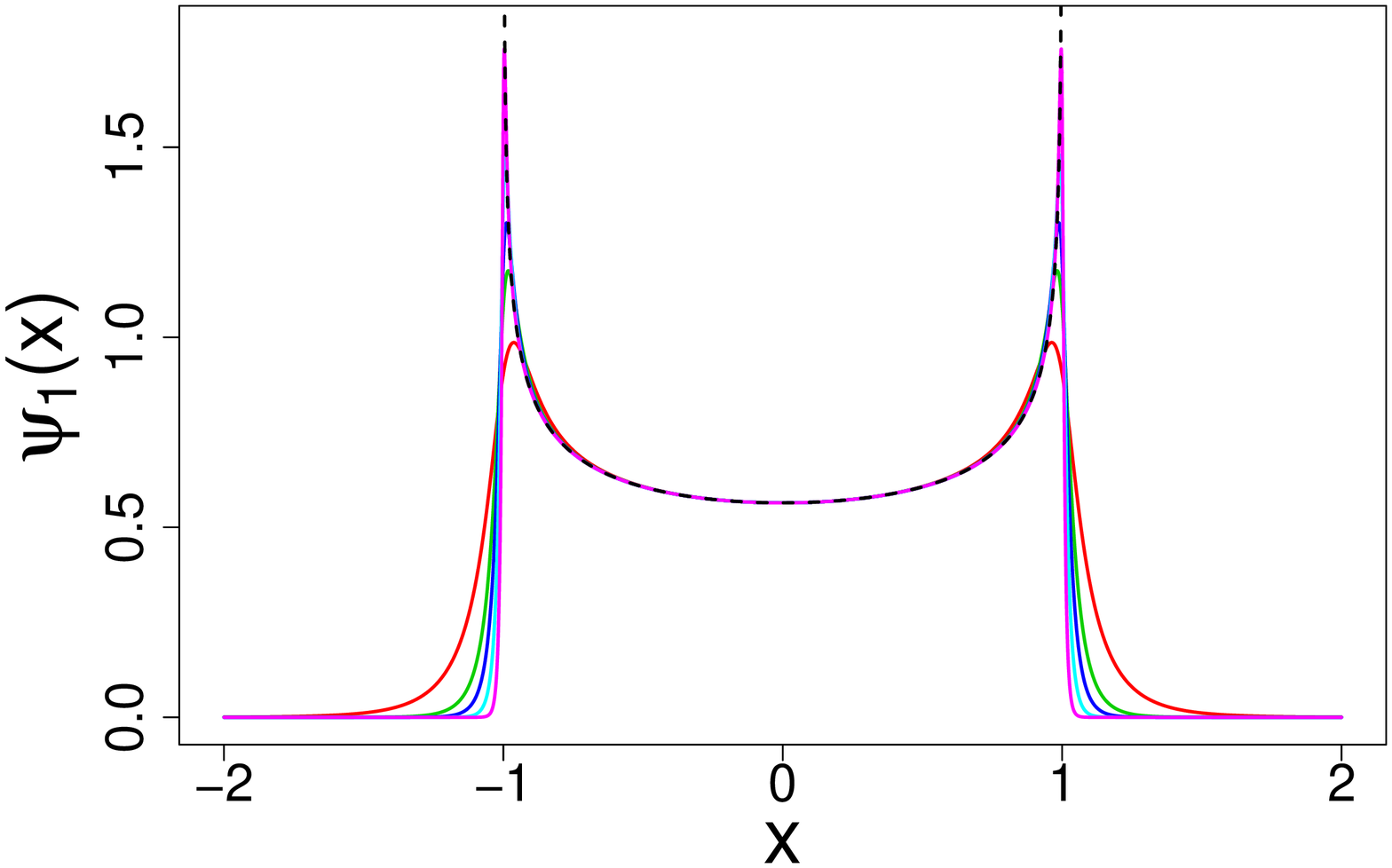}
\includegraphics[width=90mm,height=80mm]{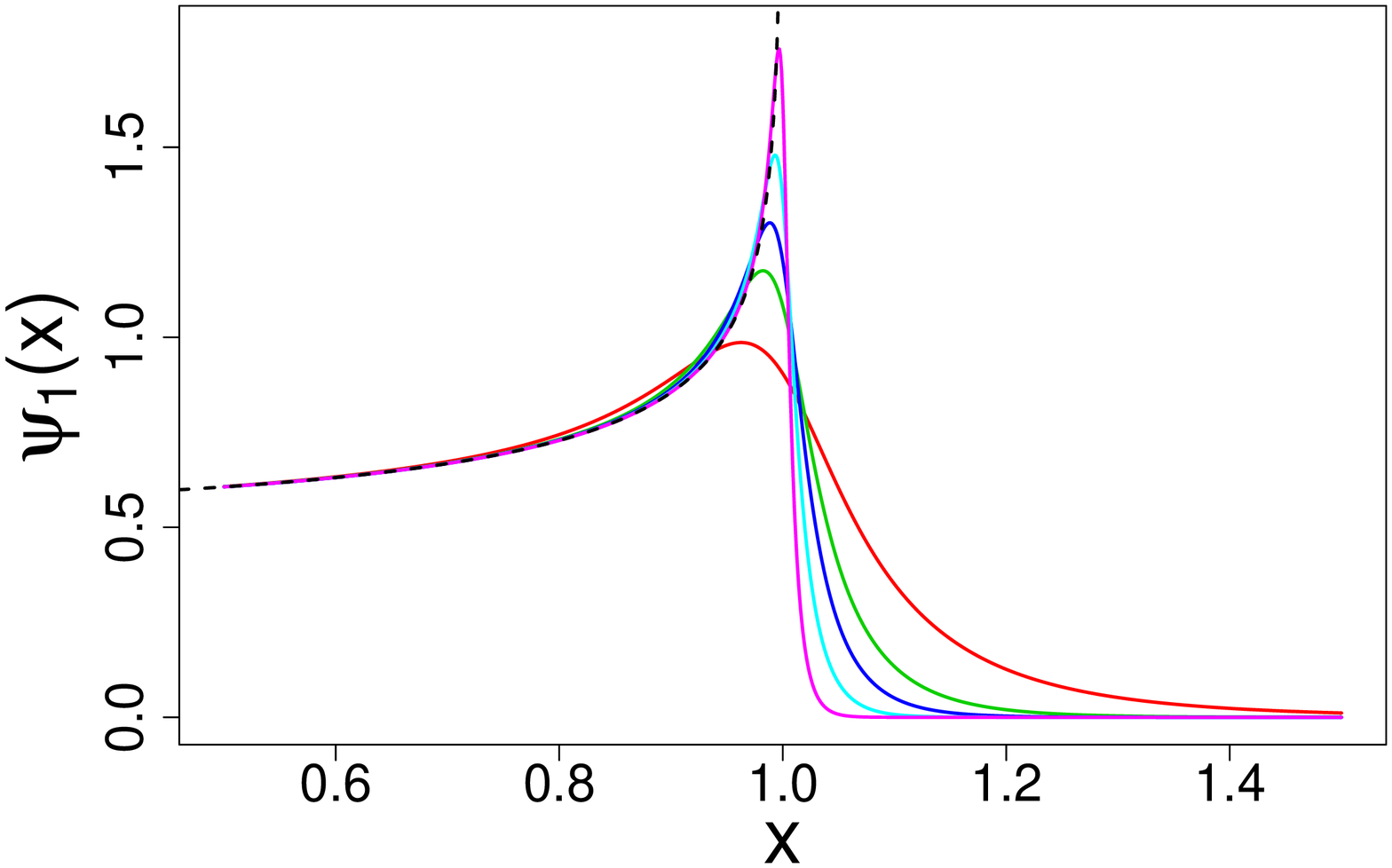}
\caption{Left panel: $\psi_1(x)\equiv  \rho _*^{1/2}(x)$ for  $m = 10, 20, 30, 50, 100$. Indices correspond to
 consecutive maxima in the growing order. Right  panel: enlarged  vicinity of   $x=1$.
  For large values of  $m$ (say $m\geq 100$),  within  the adopted graphical resolution limits,
    shapes of  superharmonic  ground states  functions  are practically indistinguishable from
     the square root of the arcsine law:  $\rho _*^{1/2}[m \to \infty ](x) \sim  (1-x^2)^{-1/4}$,
      depicted by a dotted line. Compare e.g. \cite{dubkov1,dybiec,denisov}.}
\end{center}
\end{figure}

\begin{figure}[h]
\begin{center}
\centering
\includegraphics[width=75mm,height=80mm]{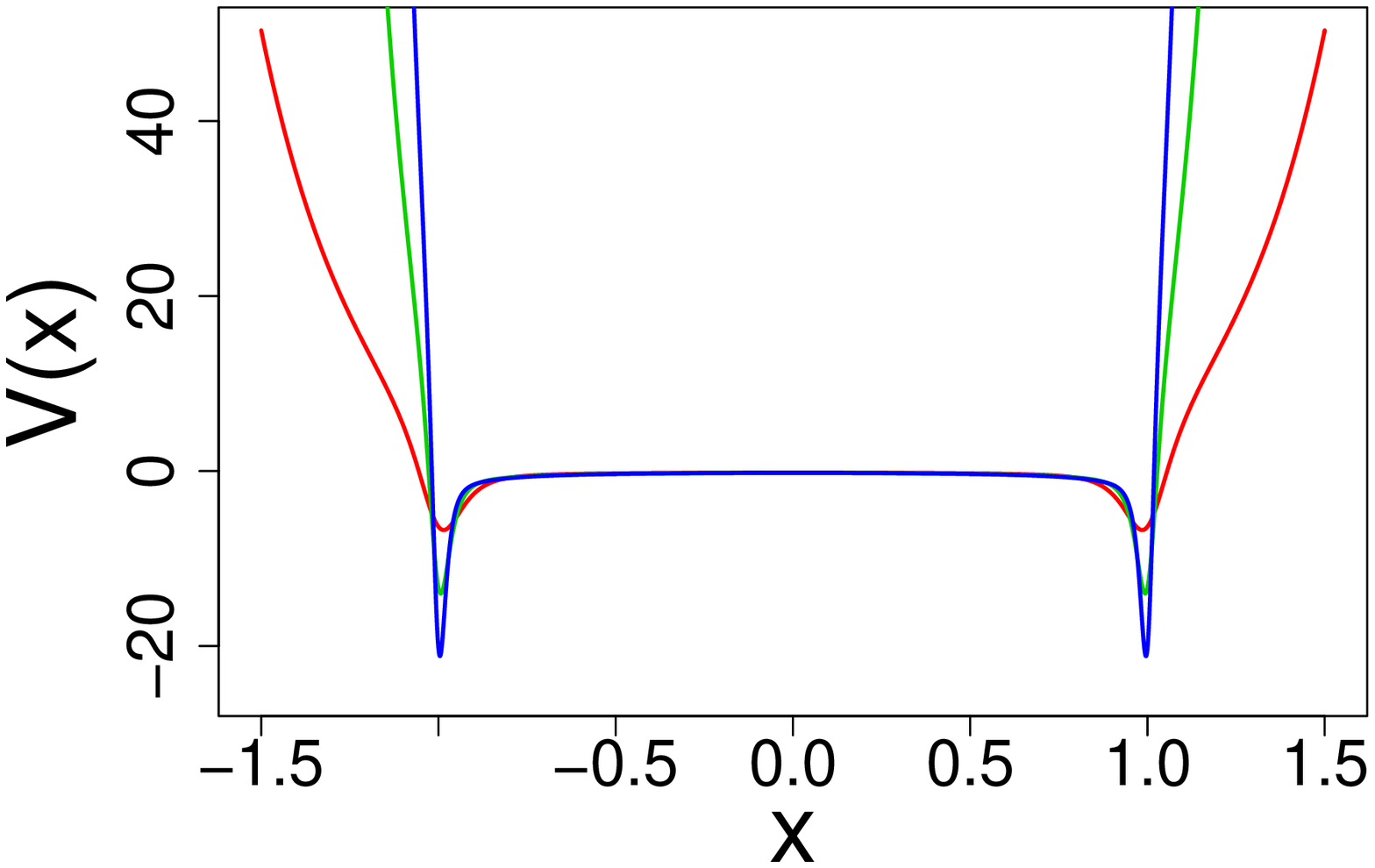}
\includegraphics[width=95mm,height=80mm]{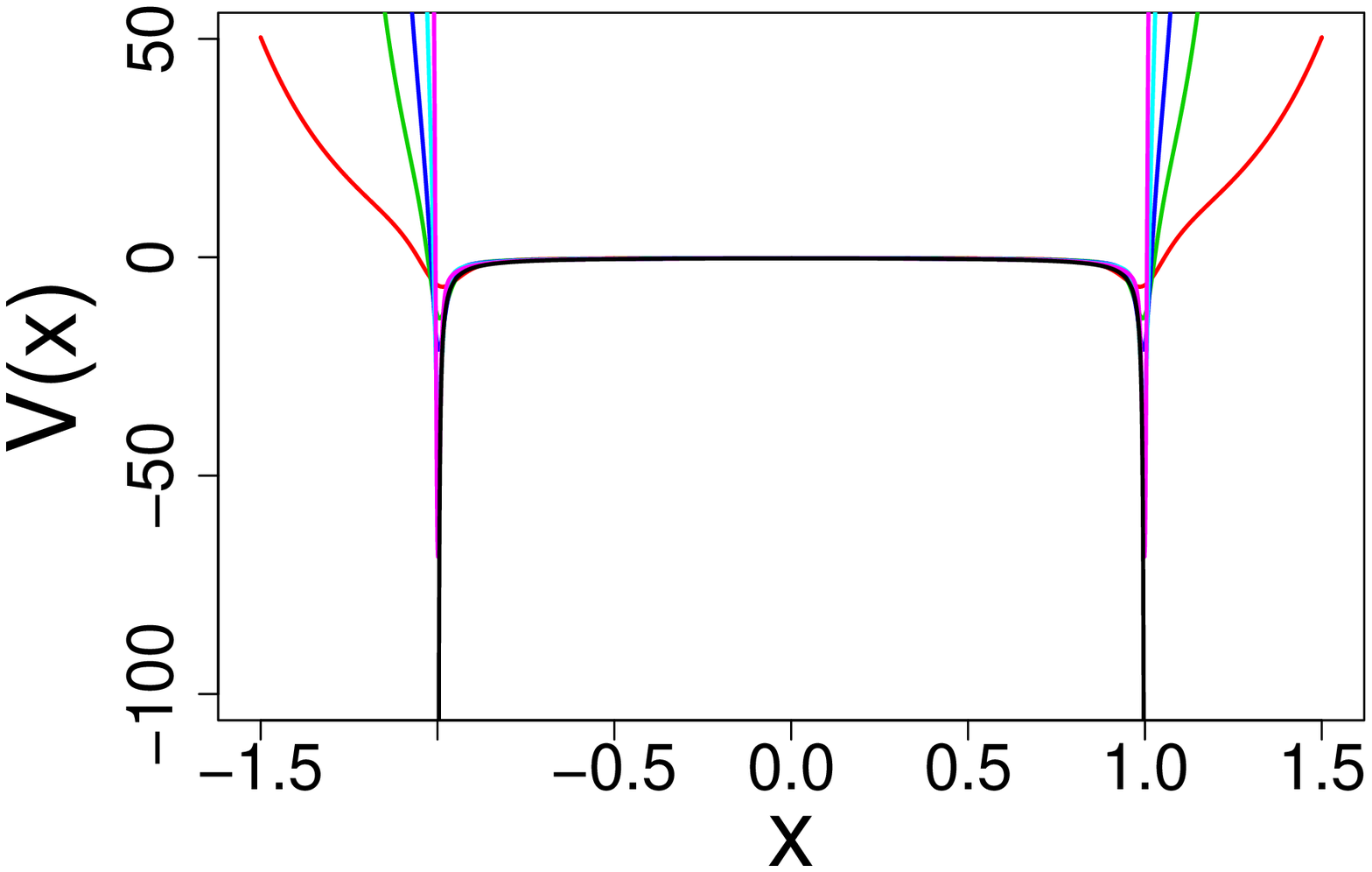}
\caption{Both panels  depict the potential ${\cal{V}}(x) = -  \rho _*^{-1/2} (x) |\Delta | ^{1/2} \rho _*^{1/2}(x)$, with $m$-labels and colors shared with Figs. 5 and 7.
Left panel:  for $m=10$ (red), $20$ (green), $30$ (blue).  Right panel: $m= 50$  (turquois) and $100$ (pink)
are additionally  inserted, compare e.g. Fig. 7.}
\end{center}
\end{figure}

\subsection{Large $m$ behavior:  boundary jeopardies.}

Let us take advantage of the rearrangement of formulas (51), (52), deduced in the Appendix of Ref.
 \cite{dubkov1}, which explicitly captures   the near-boundary behavior  in the interval   $[-1,1]\subset R$.
(The derivation is formal  under a presumption   that no obstacles come from  infinite summations).

We have the following expression for the $2m$-th pdf (the original formula refers to   the interval
 $[-L,L]$ and  has more clumsy form in view of the  presence of dimensional constants):
\be
\rho _*(x)={\frac{1}{\pi }} \exp \left\{ \sum_{l=1}^{\infty } {\frac{x^{2l}}{2l \cos [\pi l/(2m-1)]}}\right\}
\ee
in the interior  $(-1,1)$ of the interval of interest  (we note that originally \cite{dubkov1}
the statement was
"for $|x|\leq 1$", hence referred to the endpoints of the interval as  well). The formula   valid in  the exterior of  the interval (originally "for $|x|>1$") reads:
\be
\rho _*(x)={\frac{1}{\pi x^{2m} }}  \exp \left\{ \sum_{l=1}^{\infty } {\frac{1}{2l x^{2l} \cos [\pi l/(2m-1)]}}\right\}.
\ee

Let us analyze point-wise the large $m$ behavior of the definition  (53), (54) (to be considered jointly on $R$)
of the $2m$-th pdf $\rho _*(x)$.  First let us notice that (53) and (54) actually coincide at endpoints $\pm 1$
of the interval of interest, provided the series converge. This is the case for all finite values of $m$.
However $\rho _*(\pm 1)$ diverges   in the $m\to \infty $ limit,  which is  the property  of both expressions (53), (54).

Let us assume that $|x|>1$. For all finite values of $m$ the large $m$ behavior of Eq. (54) is ruled by the
 factor $1/x^{2m}$.  Hence for all $|x|>1$, $\rho _*(x)$   (and likewise $\rho _*^{1/2}(x)$ rapidly   drops
  down to zero as $m \to \infty$. Compare   e.g. Figs. 3 and 4.

 For all $|x|<1$  the infinite series in the exponent  of Eq. (53)  converge to finite values, producing function   shapes of Figs. 3 and 4, with a  visually  persuasive   $m$-dependence, showing symptoms of the     convergence to the arcsine distribution (alternatively,  its square root). This  feature  deserves a more detailed examination.

  To this end, presuming $|x|<1$, let us pass to the (formal) $m \to \infty  $ limit in the exponent of Eq. (53).  We realize that:
 \be
 \rho _*(x)_{m \to \infty } =  {\frac{1}{\pi }} \exp \left\{{\frac{1}2}
 \sum_{l=1}^{\infty } {\frac{x^{2l}}{l}}\right\}.
 \ee
 The series in the exponent  can be summed by  invoking the  Taylor expansion  of the  function
 $\ln (1-z)= -(z+ z^2/2+ z^3/3+...)$, which upon substitution $z \rightarrow  x^2$   gives rise to the
  arcsine distribution in $(-1,1)$:
 \be
 \rho _*(x)_{m \to \infty } = {\frac{1}{\pi }} \exp [-  {\frac{1}2}  \ln (1-x^2)]  = {\frac{1}{\pi \sqrt{1-x^2}}}.
 \ee

We note that the arcsine distribution has been here associated exclusively with the interior   $(-1,1)$ of the interval $[-1,1]$,
 with  endpoints   (and the whole exterior of $[-1,1]$ in $R$0  excluded from consideration. The pertinent distribution diverges as we approach  $\pm 1$.

 \begin{figure}[ht]
\begin{center}
\centering
\includegraphics[width=140mm,height=100mm]{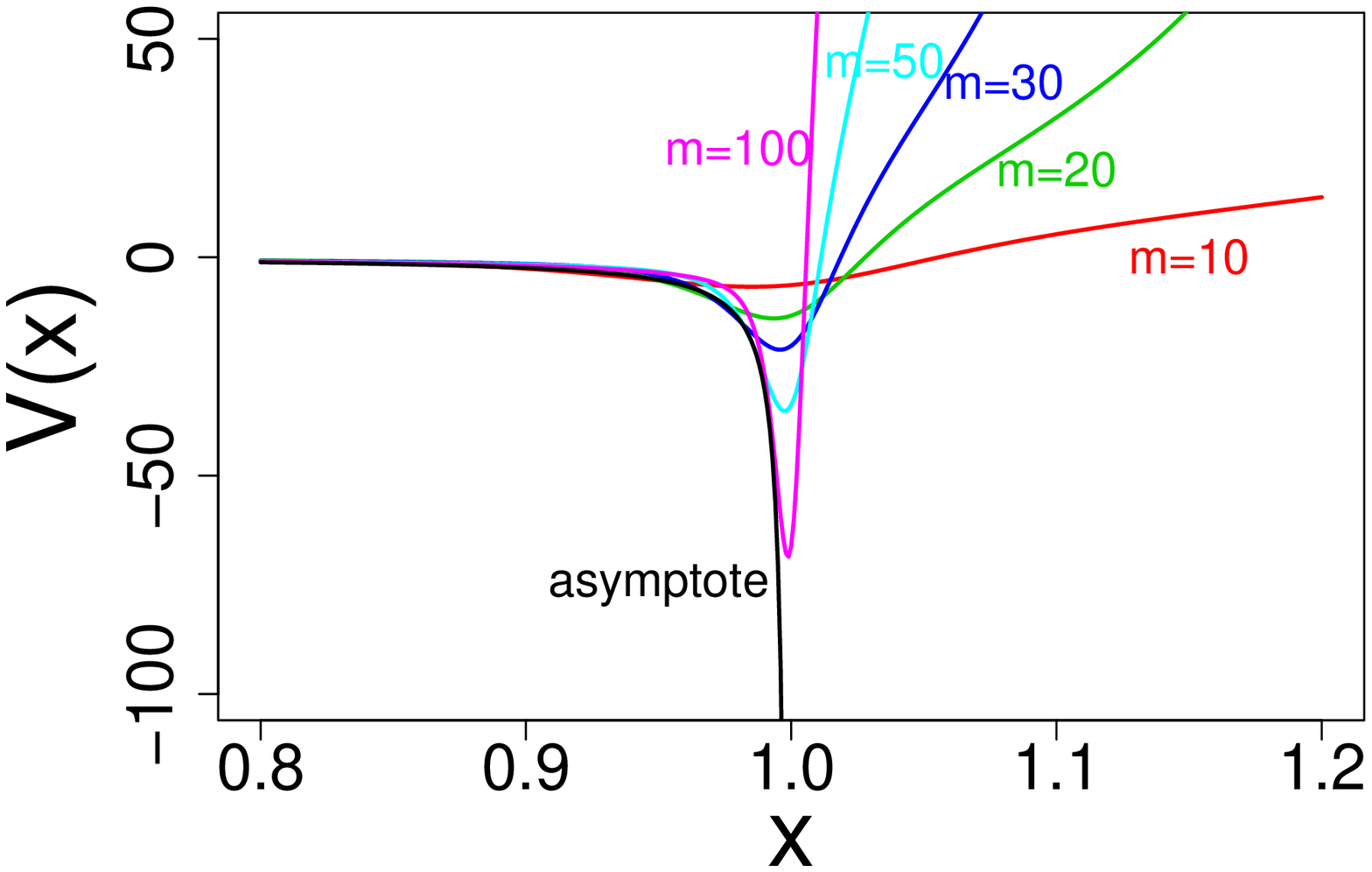}
\caption{The enlargement of the vicinity of $x=1$ in Fig. 6 clearly shows   a  surprising affinity (approximation finesse in the range ${\cal{V}}(x) \leq 0$)  set against   a   coexisting  dramatic
  difference (in the vicinity of the interval endpoints)   between  overall   shapes of  ${\cal{V}}(x)$ for finite (albeit arbitrarily large)  values of $m$ and the shape of   the  "asymptote" ${\cal{V}}(x)[m \to \infty ]$  (black), which  is  reconstructed from  $\rho _*^{1/2}(x)[m\to \infty ]$, Eq. (57).}
\end{center}
\end{figure}

 \subsection{Non-Langevin approach: Cauchy-Schr\"{o}dinger  semigroup and its equilibrium state.}

In accordance with arguments of Section II, the non-Langevin approach amounts to  the reconstruction of the dynamics   from the ground state function of a  suitable  fractional  energy operator (fractional Hamiltonian).
We have in hands the Langevin-FP induced   superharmonic  stationary densities.  These are  supposed to be
shared by the non-Langevin alternative as well.

 The   $m$-labeled ground state functions are numerically  inferred by taking
 the square root of the  $m$-th  stationary pdf:  $ \psi _1(x) = \rho _*^{1/2}(x)$  and depicted  in Fig. 5.  We note that, like in Fig. 4,
 all maxima of  superharmonic functions  (pdfs and their square roots)  {\it  are located    below} the arcsine curve  and  its square root, correspondingly.  The arcsine curves cannot be considered as envelopes  of superharmonic function  families  since, generically  in each branch,  they  have two intersection points  (hence no tangent point)  with each  superharmonic one. Nonetheless, for large values of $m$, their rough interpretation as  {\it   fapp}   envelopes  is surely admissible ("fapp" abbreviates "for all practical purposes").

To infer the Feynman-Kac (landscape) potentials ${\cal{V}}(x)$, in accordance with the discussion of Section II,  we  rely on   the numerically assisted procedure as well. Its workings (specifically on how to  handle effects of  integration cutoffs)   have  been  tested before, \cite{gar,gar1,zaba,zaba1,gar0,gar2}, see also \cite{gar5}   and  \cite{brownian}.  Given  $\psi _1(x)$, we numerically evaluate  the integral (5), while adjusted to the Cauchy case   $ - (|\Delta |^{1/2} \psi _1) (x)$, and  point-wise divide  the outcome by   $\psi _1(x)$, so arriving (point-wise again) at  the resultant ${\cal{V}}(x)$.

 The Feynman-Kac   potentials ${\cal{V}}(x)$,  inferred from a priori given   $m$-sequence   $\psi _1(x)$,  are depicted in Figs. 6 and 7.   In  the right panel of Fig.6,  and its enlargement in the vicinity of $x=1$, displayed in  Fig. 7, the black curve depicts the potential shape   ${\cal{V}} [m\to \infty ](x)$, obtained by adopting the general  reconstruction  recipe (c.f. Eq. (27)), to the square root  of the
 formal asymptotic ($m \to \infty $)  steady state  Cauchy  solution   $\rho _*(x)$, Eq. (57) and   \cite{dubkov1}.

 Following our non-Langevin reconstruction principles of Section II, we infer the Hamiltonian $\hat{H} =  |\Delta |^{1/2} + {\cal{V}}(x)$, whose ground state eigenfunction ($m\to \infty $ indication is  formal, the convergence is not uniform)
 \be
  \rho ^{1/2}_*(x)_{m \to \infty } =  {\frac{1}{\sqrt{\pi \sqrt{1-x^2}}}}
\ee
is associated with  zero eigenvalue.  To this end, we have numerically computed  the corresponding ${\cal{V}}(x)$, see e.g. Figs. 6 and 7.  Note  an excellent (within graphical resolution limits) approximation of superharmonic potential profiles  by an asymptote (57), delineated in black in Figs. 6 and 7,  in the open interval $(-1,1)$. The behavior of superharmonic potential  profiles at the close vicinity of  interval endpoints is drastically different (violent decay followed by violent growth, c.f. the case of $m=50$ and   $m=100$ in Fig. 7)   from the monotonic decay   of the curve (57)   towards minus infinity  at both endpoints $\pm 1$ of the interval.

 The black-colored curve in Figs. 6 and 7 delineates the potential, which is entirely confined  in the interval $[-1,1]$. The potential  is non-positive,   with branches rapidly  escaping  to $- \infty $ at the interval endpoints.  To the contrary, the superharmonic profiles  ultimately   escape   to $+ \infty $, while approaching $\pm 1$.

  Making a naive, but straightforward  comparison with the inverted harmonic oscillator potential, we realize that the problem refers to the scattering  phenomena in the interval $(-1,1)$.   Here, one needs an  additional information (absent in the formal definition of  the pertinent  ${\cal{V}}(x)$) that the scattering is actually limited by impenetrable walls at $\pm 1$.

 It seems that  we have  nothing to say about the well (interval)  exterior  in $R$.  However,  we shall demonstrate that  actually one cannot  disregard  the exterior   $R \setminus (-1,1)$ of the interval $(-1,1)$,  in any discussion of confined L\'{e}vy processes. It  has been the case for  Dirichlet enclosures of Section III,  and appears to be a general feature of  nonlocal  generators of L\'{e}vy flights   in "confined geometries".

\section{Meaning of   "confined geometry", confined domain" and  "infinitely deep well"  in the context of L\'{e}vy flights.}

In the statistical physics literature, the "infinitely deep  potential well"  is typically  considered  as a model of a "confined domain" with {\it impermeable boundaries}  \cite{denisov}, and thence  intuitively   associated  with reflected random motions that never leave the prescribed enclosure. This happens    both in the context of L\'{e}vy flights \cite{dubkov1}-\cite{trap},   and  in case    of the standard Brownian motion \cite{dubkov1,brownian}. This presumption is  conceptually amplified by reference to the semiclassical relative of the (quantum) infinite potential well, visualized by a mass point in uniform motion,  which is  perpetually   reflecting from  rigid walls.

However, the    "reflecting walls" association is  somewhat   misleading and stays in  plain  conflict   with our observations of Section III (see also a sample of related references). Indeed, there are other random   processes which  relax to equilibrium in a finite enclosure,  and definitely   have   not  much   in  common  with the reflection scenario, \cite{gar,gar5}.  Some of them  refer to so-called taboo processes, where  the   (originally)   absorbing  boundaries  turn over into  inaccessible ones.  The    exterior Dirichlet  boundary conditions are here implicit  and their workings were  briefly outlined in Section  III.  The corresponding   L\'{e}vy  relaxation processes  have been discussed   in Section II.

In  Refs. \cite{denisov,dubkov1}, explicitly  dealing with L\'{e}vy flights in the "infinitely deep potential well",  no link  has been established  with the  fractional Laplacian subject to any form of  Neumann boundary data (presumably nonlocal).  Hence, the notion of  a valid   generator of  the  L\'{e}vy  process  in a bounded domain with reflecting boundaries  appears to be absent (and basically remains an  open  question \cite{gar5}).\\

{\bf Remark 6:}  The issue of reflected L\'{e}vy flights  is not a novelty in the mathematical literature, and basically defined through a path-wise (Skorokhod SDE) realization, \cite{asm}.   While passing to    motion generators,  restricted forms of fractional Laplacians should enre the stage. Here, one encounters    ambiguities in the proper formulation of the Neumann-type condition.   In the present paper we leave aside a discussion  of  various  approaches to this issue and  refer to \cite{gar5,asm,valdi,abat0,warma,guan,guan1,bogdan0,bogdan}. We point out that  so called regional fractional Laplacians have been identified as generators of reflected L\'{e}vy processes in Ref. \cite{guan,guan1}, see also \cite{bogdan} for a discussion of censored L\'{e}vy flights. \\

In Ref. \cite{denisov},  by departing from   the Langevin-Fokker-Planck approach to  the study of L\'{e}vy flights in the "infinitely deep potential well", the analytic formula for   steady states has been  derived: "it is shown that L\'{e}vy flights are distributed according to the beta distribution, whose probability density becomes singular at the boundaries of the well. The origin of preferred concentration of flying objects near the boundaries in nonequilibrium systems is clarified".
(In passing, we mention that the arcsine distribution  (56)  is a member of the  above  beta  pdfs  family, \cite{dubkov1}.)

The above statement   needs to be kept under scrutiny, since the infnite well enclosure is known to  admit  Dirichlet boundaries which are inaccessible. The boundary value zero for functions in the  domain of the Dirichlet fractional Laplacian  is approached continuously (there is no  accumulation of flying objects near the boundary, they are driven away).
 An  obvious conflict  of the {\it probability accumulation}  statement,   with a plentitude of results obtained for  L\'{e}vy flights in  the   Dirichlet  well  (see Section III and references cited therein), makes   tempting  to inquire into  the roots of this   incongruence. We shall  comparatively address this issue  in below.

\subsection{How does the steady-state  distribution of L\'{e}vy  flights derive under the infinite well conditions: The argument of Ref. \cite{denisov}.}

In Ref. \cite{denisov}, the   departure point is a  formal  Langevin equation $ \dot{x}(t)= f(x(t),t) + \xi (t)$
where $f(x,t) =- \nabla U(x,t) $ is a force field, $U(x)$ an external deterministic potential, and $\xi (t)$ stands for the L\'{e}vy "white noise". Ways to handle the formal "noise" term , and derive an associated fractional Fokker-Planck equation, have been described in \cite{denisov}, see also \cite{fogedby}.
The Authors prefer to employ  the Riemann-Liouville derivatives, which  in case of the symmetric L\'{e}vy stable noise imply  a familiar \cite{fogedby} expression for the L\'{e}vy Fokker-Planck equation for a time dependent probability distribution $\rho (x,t)$:
\be
{\frac{\partial \rho (x,t)}{\partial t}} = -  {\frac{\partial [f(x,t) \rho (x,t)]}{\partial x}} + \gamma
 {\frac{\partial ^{\alpha } \rho (x,t)}{\partial |x|^{\alpha }}},
\ee
where $\gamma $ stands for the "noise" intensity parameter (to be scaled away),  while the fractional derivative conforms with our previous notation, according to: $|\Delta |^{\alpha /2} \rho (x,t) =  -  \partial ^{\alpha } \rho (x,t)/\partial |x|^{\alpha }$, $0 <\alpha <2$.

The "confined geometry" of the infinitely deep potential well    is   created  by demanding: (i) $f(x,t)=0 $ within the well, i.e.  for $x \in [-L,L]$, where $2L$ is the width   of the well, (ii) boundaries at $x=\pm L $ are impermeable, i. e. $\rho (x,t)=0$ for $|x|>L$; this restriction tacitly presumes that the term $f(x,t)\cdot  \rho (x,t)$ may be safely  discarded if  $|x|>L$, while nothing is said about what   actually  $f(x,t)$   outside the well is.

The subsequent conclusion of \cite{denisov} reads: "with these conditions the [fractional Fokker-Planck] equation  for the stationary probability density $\rho _*(x)$  reduces to"  (here, we employ the  notation of Eq. (58)):
\be
|\Delta |^{\alpha /2} \rho _*(x)  =0,
\ee
where $\rho _*(x)=0$  for $|x|>L$,  and  nothing is said about the (non)existence or specific values taken by   $\rho _*(x)$ at  boundary  points  $x= \pm L$ .

Presuming that the fractional Fokker-Planck equation  (58) and likewise  its stationary variant  (59), can be (in the least formally) rewritten in the divergence form    $\partial _t\rho (x,t)  = - \nabla  j(x,t)$ where $j(x,t)$ is interpreted as a  probability current. Accordingly (59), while presented in the time-independent form  $\nabla j(x)  =0$,  says that $j(x) = const $. At this point a boundary condition  upon the probability flow intervenes: (iii) $j(\pm L) =0$, whose consequence is $j(x)=0$ for all $x\in [-1,1]$. A hidden assumption is that $j(x)$ may be continuously interpolated up to the boundaries, while $\rho _*(x)$ may not.\\

{\bf Remark 7:} A formulation of the fractional analog of the Fick law is subtle, and likewise an inversion $\nabla ^{-1}$ of the gradient   operator is a subtle matter.  This cannot be considered  a priori granted, and  the procedure   may fail in some stability parameter $\alpha $  ranges, c.f. \cite{neel,gar7}. Then, the notion of a (fractional)  probability current cannot be  introduced  at all. \\

Conditions (i)-(iii) suggest the  functional  trial form of the sought for solution, and the subsequent computation, while restricted to  symmetric   L\'{e}vy flights   ($0<\alpha <2$)   in the well,  ends up with the  $\alpha $-family of   probability density functions in $(-1,1)$,  which  escape  to $+\infty $ while approaching the interval endpoints $\pm L$:
\be
\rho _*(x)= (2L)^{1- \alpha }\,  {\frac{\Gamma (\alpha )}{\Gamma ^2(\alpha /2)}}  (L^2 - x^2)^{\alpha /2-1}.
\ee

For the Cauchy noise $\alpha =1$,  and with  the choice $L=1$   of the interval length parameter, we arrive at  the arcsine law  in the form (56).

\subsection{ Boundary data issue.}

Our notational convention of Section I.A   gives preference to the  nonegative operator $|\Delta |^{\alpha /2}$,  while one should keep in memory that it is  $ -|\Delta |^{\alpha /2}= - (-\Delta )^{\alpha /2} $ which is a valid fractional relative   of the ordinary Laplacian  $\Delta $.      With  reference to the  normalization coefficient ${\cal{A}}_{\alpha }$ our version (c.f. Eq. (5)) is  specialized to one spatial dimension and ultimately to the Cauchy case $\alpha =1$.

 To avoid  confusion, we recall  an often employed    definition of the symmetric L\'{e}vy stable generator in $R^n$, in the integral form  which involves an evaluation of the Cauchy principal value  (p.v.).    For  a  suitable function $f(x)$, with $x\in R^n$ and $n\geq 1$, we have:
 \be
|\Delta |^{\alpha /2}f(x) =   (-\Delta)^{\alpha /2}f(x)=\mathcal{A}_{\alpha,n} \lim\limits_{\varepsilon\to 0^+}
\int\limits_{{R}^n\supset \{|y-x|>\varepsilon\}}
\frac{f(x)-f(y)}{|x-y|^{\alpha +n}}dy, \ee
where  $dy \equiv d^ny$ and  the (normalization) coefficient
\be
\mathcal{A}_{\alpha,n}=
 \frac{2^{\alpha } \Gamma ({\frac{\alpha + n}{2}})}{\pi ^{n/2}
  |\Gamma (- {\frac{\alpha }{2}})|}  =
  \frac{2^{\alpha } \alpha \Gamma ({\frac{\alpha + n}{2}})}{{\pi ^{n/2}
  \Gamma (1- \alpha /2})}
\ee
is  adjusted  to secure   the  conformity of the  integral   definition (61)  with its Fourier transformed version. The latter  actually gives  rise to the
 Fourier multiplier representation  of the fractional Laplacian, c.f.  \cite{gar7,abat},
${\cal{F}} [(- \Delta )^{\alpha /2} f](k) = |k|^{\alpha } {\cal{F}} [f](k)$.
 If the fractional operator (61) is defined on $R$, the coefficient (62) can be recast in  the form,  made explicit in Eq. (5).

Let us assume to have given a function $f(x)$,  defined on the whole of $R$,  which has the form  $f(x)=u(x)= (1-x^2)^{-1+ \alpha /2}$  for $x\in (-1,1)$ and vanishes   outside of the open interval  $(-1,1)$, e.g. $f(x)=0$ for $x\in R\setminus (-1,1)$. Thus, our function  is presumed to vanish  both  {\bf  at }   the  boundary points (endpoints) $\pm 1$ {\it and  beyond}  $[-1,1]$ as well.

  The computational outcome of Ref. \cite{dyda} reads $|\Delta |^{\alpha /2} u(x)= 0$ for all $x\in (-1,1)$.
An analogous outcome is obtained for  functions    $v(x)= x u(x)$. There holds   $|\Delta |^{\alpha /2} v(x)= 0$ as well, for all $x\in (-1,1)$. Functions that remain constant in $D=(-1,1)$ and vanish in $R\setminus D$, are valid elements of the (domain) kernel of the operator  $|\Delta |^{\alpha /2}$ as well.

 The Cauchy case refers to $\alpha =1$, and the arcsine law (56), while {\it  extended} to the whole $R$, as a function identically vanishing on the complement of $(-1,1)$ provides an example of the above introduced function $u(x)$.\\

{\bf Remark 8:}    We point out that the  computation of eigenfunctions and eigenvalues  of the (Cauchy)   fractional   Laplacian with exterior Dirichlet boundary  conditions (e.g. that in the "infinite potential well"),  c.f. Section III.C of ref. \cite{zaba1}, makes  an  explicit usage of the  assumption that the  (bounded)  eigenfunctions   $\psi (x)$ of $|\Delta |^{1/2}$  continuously approach the value  $0$, while reaching the endpoints  $\pm 1$  of   $[-1,1]$   from the its interior $(-1,1)$.   C.f. Eqs. (5)-(7) and (47) in Ref. \cite{zaba1}, where this  demand  is explicitly stated
 \be
\lim_{x\to \pm1} |\Delta |^{1/2} \psi (x) = 0,
\ee
for solutions of  the eigenvalue problem $|\Delta |^{1/2} \psi (x) = \lambda \,  \psi (x)$,  with $\lambda \geq 0$  (actually $\lambda >0$  in the  exterior  Dirichlet enclosure, \cite{kulczycki}-\cite{zaba3}). \\

 It is the    condition  (63)  which    makes  a   crucial    difference between two "infinitely deep well" cases discussed in the present paper: this described in Section III  respects (64), while  this outlined in the present Section, following Ref. \cite{denisov}, does not.   The difference is  evidenced in the boundary properties of  functions  depicted in Fig. 1  (convergence to $0$) and the divergence of   steady state  pdfs  (46) and (60), as depicted in Fig. 4. We shall come back to this point in below, by means of analytic arguments.

\subsection{Singular  $\alpha$-harmonic functions.}

Since functions (56) and (60) may be interpreted  as solutions of the fractional Laplacian eigenvalue problem with the eigenvalue zero, we   have here    a natural link with the concept of  singular $\alpha $-harmonic functions \cite{bogdan,ryznar} and  closely related   blow-up phenomena for "large solutions" of fractional elliptic equations   \cite{abat}-\cite{vazquez}.   An inspection of  Fig. 4  reveals  another link,  with the concept of  locally accurate approximations to "almost every function", which are    provided by suitable $\alpha $-harmonic ones.

In the Cauchy context, we have an explicit statement,  \cite{ryznar} concerning the concept of $\alpha $-harmonicity.  Let $\alpha <2$, and suppose that  $D$ is an {\it open unit ball} in   $R^n$. Then,  the function $f(x)= (1- |x|^2)^{\alpha /2 -1}$ for $|x|<1$   and $f(x)=0$  for $|x|\geq 1$ is $\alpha$-harmonic in $D$ and  $\lim_{x\to  Q \in \partial D} = \infty $. Here $Q$ refers to points of the boundary  $\partial D$ of $D$ (i.e. interval endpoints if $n=1$ and  $D=(-1,1)$).

 According to Refs. \cite{bogdan0,bogdan1}: (i) a function $f$  is   {\it singular}  $\alpha $-harmonic in an open set $D$ if it is $\alpha $-harmonic in $D$ and $f(x)=0$ for $x\in D^c= R^n \setminus D$;   (ii) a function $f$ is $\alpha $-harmonic in $D$ if and only if it is $C^2$ on $D$ and   $|\Delta |^{\alpha /2} f(x)=0$ for all $x \in D$.

It is (ii), which directly refers to our previous discussion.  We emphasize that for  (ii) to hold true, the function $f(x)$ must be defined on the whole of $R^n$.  The values of $f(x)$  on $D^c$  are indispensable  for this property   and  {\it must not}   be
  disregarded (or ignored). This reflects the fact that the fractional Laplacian is a nonlocal operator and without special precautions \cite{bogdan,gar5} there is no way to eliminate  a direct influence (e.g. jumps)  between distant points $x$ and $y$ in the domain of $f$.

On the other hand, the notion of $\alpha $- harmonicity  can be introduced in the purely probabilistic  lore,   with direct reference  to L\'{e}vy flights, thus  providing hints toward computer-assisted path-wise procedures, yielding the singular $\alpha $-harmonic functions as would-be   steady states of L\'{e}vy flights in the  (appropriately defined)  "infinite potential well", \cite{dubkov1,dybiec,denisov}.

 To this  (probabilistic/stochastic)  end, c.f. \cite{bogdan0,bogdan1}, we employ the notion of    the   first exit  time from $A \subset R^n$  (alternatively, first  entrance time to $A^c = R^n \setminus A$)  of  the  isotropic  $\alpha $-stable   L\'{e}vy process  $X_t$.    Given the Borel set $A$, we define $\tau  _A=  \inf \{t\geq 0: X_t \in A^c\}$   as  the first exit time from $A$.   For a bounded set $A$, we have $\tau _A< \infty  \, \,  a.s.$.   We define  a local expectation value
  \be
  u(x)=  E^xu(X_{\tau _A}) = E^x[u(X_{\tau _A}); \, \tau _A < \infty ],
 \ee
   interpreted as an average   taken at  random (exit/entrance) time $\tau _A$ values, with respect to the process $X_t$ started in $x$ at $t=0$, with values $X_t=y \in A$.

For a Borel measurable function $u \geq 0$ on $R^n$,  we say that:\\
  (i)  $u(x)$  is  {\it regular }  $\alpha $-harmonic in an open set  $A\subset R^n$, if   $u(x) = E^x[u(X_{\tau A})] <\infty $, $x\in A$; \\
  (ii)   $u(x)$ is $\alpha $-harmonic in $A$, if
  for every bounded open set $B$ with the closure $\overline{B}$ contained in $A$  we have $u(x) = E^x[u(X_{\tau B})]  < \infty $, $x\in B$;  \\
   (iii) $u(x)$  is {\it singular}  $\alpha $-harmonic  in $A$, if $u(x)$ actually is $\alpha $-harmonic  in $A$ and  $u(x) = 0$ for all $x\in A^c$.\\

Accordingly, the "steady state functions" (56) and (60), while  interpreted as  valid solutions of Eqs. (63) and (59) respectively (and thus  complemented by the exterior boundary condition), are  examples of   singular $\alpha $-harmonic functions in $D=(-1,1)$.

We note that a   visual  inspection of both panels in Fig. 4 clearly indicates that, while in $(-1,1)$,  the singular $\alpha $-harmonic function $\rho _*(x)$ may be considered as a perfect approximation of large $m$ superharmonic pdfs.  (This property stays in conformity with recent Ref. \cite{valdi1}, according to which "all functions are locally $\alpha $-harmonic up to a small error".)

In the exterior of $[-1,1]$ i.e. for $x>1$,  the pertinent  pdfs, in the large $m$  regime,   rapidly decay to zero. The behavior (with $m \to \infty $) of these pdfs is subtle: we have (i) growth to $\infty $ for $|x|\leq 1, |x| \uparrow 1$, and (ii) decay to zero for $|x|\geq 1, |x| \downarrow 1$.

\subsection{Domain intricacies and the relevance of  exterior contributions.}

A formal statement of  the  exterior Dirichlet boundary data for  the fractional operator (negative  fractional Laplacian)
 $|\Delta |^{\alpha /2}$   may be condensed in the notion of the  elliptic problem, \cite{warma}: $
 |\Delta |^{\alpha /2} u = f$  in  $\Omega $ and $u = 0$  in $R^n\setminus  \Omega $
where $\Omega \subset R^n$ is an arbitrary bounded open set.  Actually, this is  a departure point for
 the study of   the  eigenvalue  problem  $ |\Delta |^{\alpha /2} u =  \lambda  u$, provided   $|\Delta |^{\alpha /2}$  has a realization  in  $L^2(\Omega )$,   e.g.   both $u$ and $f$  are  elements of   $L^2(\Omega )$.  The eigenvalues  $\lambda $   are  known to be positive, $\lambda >0$ , c.f. \cite{kulczycki,kwasnicki,zaba,zaba1}.

The Cauchy  ($\alpha =1$)  version of the  pertinent  spectral problem,    under  exterior  Dirichlet  boundary  data,  has been  briefly  summarized   in  Section III, with the notational replacement of $\Omega $ by $D$.

Right at this point we emphasize, that the singular $\alpha $-harmonic functions (56) and (60) formally
 correspond  to the eigenvalue  zero of the fractional Laplacian, while considered   in $L(\Omega )$, e.g.  {\it not}  in $L^2(\Omega )$.   Note that  to employ the framework of Section II, we were forced to introduce  the square root of the arcsine pdf, (57),  to deal with  the  $L^2(\Omega )$ (actually  $L^2(D)$)  setting.

Our further analysis pertains to the Cauchy case.   For a while  we disregard the  domain  issues, i.e.  $L$ versus $L^2$, and/or  involved  Sobolev spaces, \cite{warma},   and formally  address  the existence of  solutions of   the fractional  equation   $|\Delta |^{1/2} \rho _*(x)=0$  in  $D =(-1,1)$,  with exterior Dirichlet boundary  data  in   $R \setminus D$, c.f. (59).    We know that not only positive solutions  are admitted, \cite{dyda}, and that  arbitrary   constants do this  job as well.

We are interested in  an explicit  justification of  the existence of positive solutions with the blow-up at the boundaries $\pm 1$ of $D^c= [-1,1]$.   In below  we shall   analytically   demonstrate that the arcsine  pdf (56)    is an   example of a    fully-fledged singular $\alpha =1$-harmonic function in $D=(-1,1)$ and  indicate: (i)  an exterior  to $(-1,1)$ input to the solution, and (ii)   the role (acceptance or  abandonment)  of  the "continuity up to the boundary" condition (64).

We depart form the  integral  definition (4) of the fractional operator $|\Delta |^{1/2}$, where ${\cal{A}}_1= 1/\pi $.   Let us tentatively consider  the action of  $|\Delta |^{1/2}$  on $C_0^{\infty }(R)$ functions $\psi (x)$   supported in $D=(-1,1)$,   i.e.  such that  $\psi (x) = 0$ for all $x \in R\setminus D$.  We have  ($(p.v.)$ means Cauchy principal value):
\begin{equation} \label{stu1}
|\Delta |_D^{1/2}\psi(x)=-\frac{1}{\pi}  (p.v.) \int_R\frac{\psi(x+y)-\psi(x)}{y^2}dy.
\end{equation}
Given $x\in (-1,1)$,  we realize that
$\psi(x+y)$ does not  vanish   identically if $ x+y\in (-1,1)$   i.e.  for $ -1-x < y < 1-x$.
Hence,  the integration (\ref{stu1}) can be
simplified  by  decomposing  $R$ into $(-\infty <y \leq  -1-x) \cup (-1-x<y<1-x) \cup (1-x \leq y <\infty )$.  Therefore, we end up with a {\it restricted} fractional operator:
\begin{eqnarray}
|\Delta |_D^{1/2}\psi=-\frac{1}{\pi} \left[-\psi(x)\left(\int_{-\infty}^{-1-x}\frac{dy}{y^2}+\int_{1-x}^\infty\frac{dy}{y^2}\right)+
\int_{-1-x}^{1-x}\frac{\psi(x+y)-\psi(x)}{y^2}dy\right]=\nonumber \\
=\frac{2}{\pi}\frac{\psi(x)}{1-x^2} + \frac{1}{\pi}\int_{-1-x}^{1-x}\frac{\psi(x)-\psi(x+y)}{y^2}dy, \label{stu2}
\end{eqnarray}
where the second integral  should be understood as the  Cauchy principal value  with respect to $0$, i.e. $
\int_{-1-x}^{1-x}=\lim_{\varepsilon \to 0} \left[\int_{-1-x}^{-\varepsilon} +\int_{-\varepsilon}^{1-x}\right]$.

We point out that the first term on  the right-hand-side of  Eq. (66) includes an outcome of the  integration over $R\setminus D $,  i.e. an input   exterior to   $D =(-1,1)$  proper.
It is instructive to notice that the  change the integration variable  $y=t-x$ in the second term of  Eq. (66) gives rise to
 \be \label{int}
 |\Delta |_D^{1/2} \psi (x)
=\frac{2}{\pi}\frac{\psi(x)}{1-x^2}+\frac{1}{\pi}\int_{-1}^{1}\frac{\psi(x)-\psi(t)}{(t-x)^2}dt,
\ee
 where   the $R\setminus D $ and $D$ contributions  are now
clearly isolated, albeit the ultimate overall  $x$-dependence refers to $x\in D$ only.   The Cauchy principal value of the integral in Eq. (\ref{int})  is no longer evaluated with respect to $0$, but with respect to $x$.

 The integral expression in Eq. (\ref{int}) which is restricted
 to $t\in (-1,1)$, and   $x\in (-1,1)$,  is  the  Cauchy version  of the   so-called {\it regional}
  fractional  Laplacian in  $D^c =[-1,1]$, \cite{bogdan,warma,gar5,zaba1}.

Our further discussion will be based on the  decomposition  (66), which has been used in Ref.  \cite{zaba1},
  for the    computer-assisted   shape  analysis of nonolocally induced fractional  (Cauchy) bound states in the infnite well.  See e.g.  Section III of the
    present paper, and the approximate formula (49) for the ground state function, which  reproduces the
    $\sim \sqrt{1-x^2}$ decay to $0$, while approaching the boundary points $\pm 1$ of the interval
    $(-1,1)$. This, in conformity  with    the condition (63), and at variance with the  boundary  blow-up property of (56) and (60).

\subsection{Explicit evaluation of the   singular  $(\alpha =1)$-harmonic function  in  $(-1,1)$.}

In Refs. \cite{zaba,zaba1}, we have assumed that any  even  eigenfunction  of the Dirichlet fractional Laplacian $|\Delta |^{1/2}_D$ (given by Eq. (66),   restricted by the  exterior Dirichlet boundary  data  to $D$, and  additionally by  the  local boundary   condition (63)), should be sought for by  analyzing  convergence features of  consecutive  polynomial    approximations  of $2N, N\to \infty $ degree,  in  terms of power series expansion (here given up to the normalization coefficient):
\be  \label{g}
\psi(x)= \sqrt{1-x^2}\sum_{k=0}^N \alpha_{2n}x^{2n},\qquad \alpha_0=1.
\ee
 In Ref. \cite{zaba1}  our major task has been  to determine  expansion   coefficients $\alpha_{2n}$, for sufficiently long series expansion  (we have computationally reached $2N= 500$).

   Given the definition (66) of $|\Delta |_D^{1/2}$ restricted to $\psi $'s with support in $D$.    Let us  formally  proceed with its integral part,  here denoted
\be
I_D\psi(x)=\frac{1}{\pi}\int\limits_{-x-1}^{-x+1}\frac{\psi(x)-\psi(x+z)}{z^2}dz.
\ee
Consider  the (formal)    action of $I_D$ upon   functions  of the form $\psi (x) = x^{2n} \sqrt{1-x^2}$.   We get (compare e.g. \cite{zaba1}):
\be  \label{trial}
I_Dx^{2n}\sqrt{1-x^2}=-\frac{2}{\pi}\frac{x^{2n}\sqrt{1-x^2}}{1-x^2}+(c_{2n}+3c_{2n-2}x^2+5c_{2n-4}+\ldots+(2n+1)c_0x^{2n}),
\ee
where  $c_{2k}$  are  expansion coefficients of the Taylor series for $\sqrt{1-x^2}$:
\be
\sqrt{1-x^2}=\sum_{k=0}^{\infty}c_{2k}x^{2k}=\sum_{k=0}^{\infty}\frac{(2k)!}{(1-2k)(k!)^24^k}x^{2k}.
\ee

We note that
\be
 {\frac{1}{\sqrt{1-x^2}}} =    (1+x^2 + x^4+...) \sqrt{1-x^2}
\ee
where we recognize a factor which is a sum  of  a geometric progression  with the ratio $x^2, |x|<1$.  This allows to evaluate term after term (presuming suitable convergence properties of the series) the expression:
\be
|\Delta |^{1/2}_D  {\frac{1}{\sqrt{1-x^2}}} =  \sum_{n=0}^{\infty }  |\Delta |^{1/2}_D  [x^{2n} \sqrt{1-x^2}]=          \sum_{n=0}^{\infty  } c_{2n} + 3x^2  \sum_{n=0}^{\infty  } c_{2n} + 5x^4   \sum_{n=0}^{\infty  } c_{2n}  +... = 0.
\ee
Here, we note that the function $\sqrt{1-x^2}$ is defined on $[-1,1]$ and takes the value  $0$ at $x= \pm 1$.  Accordingly,  $\sum_{k=0}^{\infty } c_{2k} = 0 $ and therefore there holds the expected result
$|\Delta |^{1/2}_D  (1-x^2)^{-1/2} = 0$.

We emphasize that  in view of (70), the exterior (by origin)  term in (66) and (67)  is cancelled by the intrinsic (to $D$)  counterterm $ - 2\psi(x)/\pi (1-x^2)$.

{ \bf Remark 9:} We point out that potentially troublesome issues of the interchange of infnite summations and integrals have been bypassed. To facilitate the passage from (70) to (73), let us indicate  that: $|\Delta |^{1/2}_D \sqrt{1-x^2} =1$,  $|\Delta |^{1/2}_D x^2 \sqrt{1-x^2} =3x^2- 1/2$,
$|\Delta |^{1/2}_D x^4 \sqrt{1-x^2} = 5x^4 - 3x^2/2 - 1/8$,  $|\Delta |^{1/2}_D x^6 \sqrt{1-x^2} = 7x^6- 5x^4/2 - 3x^2/8 - 1/16$ and so on. Summation of these expressions, paralleled by collecting together terms standing at consecutive powers of $x^{2n}, n=0,1,2,...$, gives rise to   geometric progressions: $x^0(1-1/2,-1/8 -1/16-...)$, $3x^2(1-1/2-1/8-...)$, $5x^4(1-1/2-1/8-...)$ etc.

\section{Conclusions:  Path-wise justification  attempts for the relaxation process in the "confined domain".}

As a brief  introduction to subsequent comments, we list simple (Monte Carlo)  updating  scenarios, which are supposed to to mimic  the  random path reflection in the two barrier  problem (e.g. interval $(-L,L)$ or  the  infinite well set on this interval), \cite{dybiec}.\\

\noindent
(I)Reversal (wrapping): A trajectory that ends at $x<-L$ is wrapped  around the left boundary : $x\to -L +|x+L|$.\\
(II) Stopping:  A trajectory that aims  to  cross  $-L$ is paused (stopped) at $-L+ \epsilon $, where $\epsilon >0$ is fixed and  small. The point  $-L+ \epsilon $ is a  starting point for the next jump (next in terms of the simulation procedure/time). The barrier is inaccessible to the trajectory. \\
(III) Superharmonic confinement: The Langevin-type equation with a superharmonic force term $-\nabla U= - x^{2m-1}/L^{2m}$, $2m\gg 2$ is used to simulate the L\'{e}vy motion.\\

A particular property of  confined L\'{e}vy flights,  we have discussed throughout the present paper,  is the accumulation  (ultimately  interpreted as a blow-up)  of "steady state" (or equilibrium)   probability density functions (and thence probability) near the boundaries of the confining potential well, c.f.   Figs. 3  to 7 and  (56), (60).  Such phenomena have been reported  in  computer studies of anomalous diffusions and  specifically the   fractional Brownian motion, \cite{klafter}-\cite{vojta2}. In the computation,  traditional reflection-from-the-barrier   path-wise recipes ({\it wrapping scenario}) were adopted, in conjunction with steep potential well  Langevin models.

Leaving aside the case of the  fractional Brownian motion and coming back to the L\'{e}vy flights issue, we point out that   the path-wise search for a consistent implementation of the   reflecting boundary data  and the reflection event proper,  has been carried out in Refs. \cite{dybiec0,dybiec,dubkov1}.  Probability density functions were obtained by means of numerical path-wise (Monte Carlo) simulations,  based on  the  Langevin-type equation  with the fractional ($\alpha $-stable) "white noise" term.   In fact,  stochastic differential equations were numerically integrated by applying the Euler-Maruyama-Ito method, \cite{weron}. Large numbers of sample trajectories of involved random variables  $X(t)$  were generated, which  enabled an approximate  reconstruction of the pdf $\rho (x,t)$ at a chosen (large) simulation  time instant $t$.  A stabilization of outcomes  (pdf's shapes)  for time instants  $t$  large enough, has been interpreted  as a symptom of stationarity of the   asymptotic stationary pdfs $\rho _*(x)$.

In particular, for the  superharmonic confinement (case (III))  of  $\alpha $-stable  L\'{e}vy flights, the accumulation near the endpoints of the interval $[1,1]$ has been convincingly confirmed. The  interpretation, Ref. \cite{dybiec}    of the binding potential $U(x)= x^{2m}/2m$, $m \gg 1$, has been literally  coined as  that of a  model of the {\it reflecting boundary}. Exemplary simulations were executed for $2m=800$.  (Other models of  would-be  {\it reflecting} boundaries are  present in the literature as well, see e.g. \cite{vojta2}.)

 On the available graphical resolution level, a high approximation accuracy of  singular $\alpha $-stable harmonic functions has been achieved, by  means of the    {\it stopping}  scenario (II) for random motions    between two impenetrable barriers, \cite{dubkov1,dybiec}.

  A serious conceptual obstacle should be mentioned. Namely, if  for L\'{e}vy flights in the infinite well,   we  adopt the wrapping (reflection)  scenario  (I)  at the  barriers,  then irrespective of the stability index $0< \alpha <2$, the estimated  (asymptotic)  trajectory statistics corresponds to the   {\it uniform} distribution $1/2$ in $(-1,1)$, c.f. \cite{dybiec}, see e.g. also \cite{gar5,brownian}.

On the other hand,   none of the adopted path-wise reflection  scenarios has been ever   tested  (as useful or useless) in the mathematical research on   reflected L\'{e}vy flights, developed sofar, \cite{asm}-\cite{bogdan}.  Interestingly,  main  efforts were concentrated on the formulation of the fractional  (basically nonlocal)   analogue of the   Neumann condition  (as opposed to the exterior Dirichlet one).   However,  so constrained  jump-type process appear  not to be  confined in  the interior of the well, but  in principle may reach   exterior (beyond the barrier) locations. The "reflection" is mimicked by the instantaneous  return i.e. the   jump  back  to the well interior,  with a prescribed jump intensity,  \cite{valdi,warma}. This is plainly inconsistent with the   barrier  "impenetrability" notion of Ref. \cite{denisov}.

 In the mathematically oriented literature, the reflected L\'{e}vy process is often invoked on  a fairly abstract level of analysis, with no reference to explicit path-wise motion scenarios. With reference to the semigroup lore,  {\it regional} fractional  Laplacians  have been  deduced  as  legitimate generator of the reflecting L\'{e}vy process in the bounded domain, \cite{guan,guan1} and  \cite{bogdan,gar5}. In  principle (although it is not the must), the boundaries, e.g. the  interval endpoints may be reached by the  reflecting  process, but for a suitable subclass of processes  the barrier may  happen to  be inaccessible. Apart from the wealth of sophisticated arguments,   no  detailed path-wise analysis, nor analytic (spectral) properties of the pertinent reflected stochastic  process  are available in the literature.\\

Let us briefly  summarize our findings:(i) "steady state" pdfs  (56), (60) cannot be justifiably associated with the concept of  reflected L\'{e}vy flights, whose  mathematically rigorous theory is in existence, \cite{asm}-\cite{bogdan}; (ii) at variance with  superharmonically bound L\'{e}vy flights, no relaxation process has been ever found, with  the "steady state"(56) (or (60), in general) in its large time asymptotic; told otherwise, no thermalization process is known that would relax to a singular $\alpha $-harmonic function.  These topics need a deeper  analysis.\\

{\bf Acknowledgement:}   (P. G.)  would like to express his gratitude to Professors  B. Dybiec and J. L\H{o}rinczi for  explanatory correspondence on their own work.

\end{document}